\documentclass[twocolumn, apl, amsmath, amssymb, superscriptaddress]{revtex4-2}
\usepackage{epsf}      
\usepackage{graphicx}
\usepackage{color}
\usepackage{soul}
\usepackage{gensymb}
\usepackage{sidecap}
\usepackage{amsmath}
\usepackage{mathtools}

\begin{document}
\title{Disentangling octahedral distortion and symmetry breaking in the ordered double perovskite SrLaCoNbO$_6$}

\author{Ajay Kumar}
\email{ajay1@ameslab.gov}
\affiliation{Ames National Laboratory, U.S. Department of Energy, Iowa State University, Ames, Iowa 50011, USA}
\affiliation{Department of Physics, Indian Institute of Technology Delhi, Hauz Khas, New Delhi-110016, India}

\author{Clemens Ulrich}
\affiliation{School of Physics, The University of New South Wales (UNSW), Kensington, 2052 New South Wales, Sydney, Australia}

\author{Yaroslav Mudryk}
\affiliation{Ames National Laboratory, U.S. Department of Energy, Iowa State University, Ames, Iowa 50011, USA}

\author{Rajendra S. Dhaka}
\affiliation{Department of Physics, Indian Institute of Technology Delhi, Hauz Khas, New Delhi-110016, India}

\date{\today}

\begin{abstract}

We report a comprehensive high temperature structural investigation of the B-site ordered double perovskite SrLaCoNbO$_6$ using combined Raman spectroscopy and x-ray diffraction (XRD) over the range 25--900~$^\circ$C. Two distinct temperature-driven anomalies are identified at $T_\mathrm{OD} = 300\pm 25~^\circ$C and $T_\mathrm{PT} = 750\pm 25~^\circ$C through pronounced renormalizations in the phonon frequency and linewidth, accompanied by the evolution and suppression of characteristic XRD reflections. The low-temperature anomaly at $T_\mathrm{OD}$ is associated with an isostructural octahedral distortion within the monoclinic $P2_1/n$ phase. At $T_\mathrm{OD}$, the radial and angular distortion parameters of the CoO$_6$ and NbO$_6$ octahedra decrease to near-zero values before reversing upon further heating due to an inversion of the in-plane Co--O and Nb--O bond-length hierarchy. The high-temperature anomaly at $T_\mathrm{PT}$ corresponds to a symmetry-breaking structural phase transition from $P2_1/n$ to the monoclinic $I2/m$ phase, confirmed by group-theoretical analysis and Rietveld refinement. These results demonstrate that integrating temperature-dependent Raman spectroscopy with laboratory-based XRD, complemented by quantitative octahedral distortion analysis, provides a robust approach to disentangle isostructural octahedral distortions from symmetry-breaking phase transitions in complex double perovskite oxides without recourse to synchrotron or neutron diffraction facilities.

\end{abstract}

\maketitle

\section{\noindent ~Introduction}

Double perovskite oxides with the general formula $AA'BB'$O$_6$ have attracted extraordinary scientific interest over the past more than five decades owing to their remarkable structural versatility and the wealth of functional properties, emerging from the interplay among their lattice, spin, electronic, and orbital degrees of freedom~\cite{Vasala_PSSC_15, Vries_PRL_10, Iwahara_PRB_18}. By selectively substituting both A- and B-site cations, one can independently tune the ionic radii, crystal structure, valence states, transport mechanism, and magnetic interactions, providing an unparalleled platform for engineering materials with targeted properties ranging from colossal magnetoresistance and half-metallic ferromagnetism to multiferroicity, superconductivity, and quantum-spin-liquid behavior~\cite{Kobayashi_Nature_98, Erten_PRL_11, Harshman_PRB_03, Kumar_PRB_10, Singh_PRB_21}. Central to the emergent physics in these compounds is the spatial ordering of the $B$- and $B'$-site cations, which plays a crucial role in determining the octahedral distortion, crystallographic symmetry, magnetic exchange interactions, and ultimately the magnetic ground state ~\cite{King_JMC_10, Jung_PRB_07, Bos_PRB2_04, Meneghini_PRL_09, Kumar_PRB_24}.

Beyond their fundamental interest, perovskite and double-perovskite oxides have also attracted considerable attention for high-temperature energy-conversion and energy-storage technologies, including solid oxide fuel cells (SOFCs), oxygen-permeation membranes, and heterogeneous catalysis \cite{Huang_Science_06, Tsvetkov_NatMater_16, Huang_CM_09, Grimaud_NC_13}. The technological importance of these materials arises from the remarkable flexibility of the BO$_6$ octahedral framework, including B--O bond lengths, B--O--B bond angles, octahedral tilting, and crystallographic symmetry, which plays a critical role in governing oxygen-vacancy formation, oxygen-ion migration, electronic transport, thermal expansion, and catalytic activity \cite{Grimaud_NC_13, Kumar_JAP_20, Vasala_PSSC_15, Li_NatCommun_21, Kumar_JVSTA_23, Woodward_ACB_97a, Woodward_ACB_97b}. Since SOFCs operate at elevated temperatures (600--1000$^\circ$C), temperature-induced modifications of the BO$_6$ octahedra can alter transition-metal–oxygen bonding and thereby influence orbital overlap, electronic bandwidth, defect chemistry, and chemical stability \cite{Tarancon_JMC_07, Dong_JMCA_14, Kala_JMCA_26}. Therefore, understanding the evolution of these structural motifs at high temperatures is essential for establishing crystal-chemistry–property relationships and guiding the design of advanced materials for energy-conversion applications.

With increasing temperature, thermal energy progressively reduces the octahedral tilting and rotation, driving crystal structures toward higher-symmetry phases \cite{Zabalo_JAC_13, Faik_JMS_09, Gateshki_JPCM_03}.  For example, SrLa$M$RuO$_6$ with $M$ = Mg, Ni, Cu, and Zn undergoes a $P2_1/n \rightarrow R\bar{3}$ transition at approximately 500, 670, 730, and 800~K, respectively~\cite{Zabalo_JAC_13, Gateshki_MRB_03}. In contrast, Sr$_2M$WO$_6$ ($M$ = Cd, Ca, Zn, and Co) exhibits a sequence of $P2_1/n \rightarrow I4/m \rightarrow Fm\bar{3}m$ phase transitions upon heating \cite{Gateshki_JPCM_03, Faik_JMS_09}. Intermediate $I2/m$ phases have likewise been reported in several ordered double perovskites, including Sr$_2$ScSbO$_6$, Ca$_2$ScSbO$_6$, Sr$_2$YTaO$_6$, and Sr$_2$GdRuO$_6$, prior to the emergence of higher-symmetry structures~\cite{Faik_JSSC_12,Zhou_JSSC_10,Triana_MRB_11}. However, disentangling genuine symmetry-breaking phase transitions from symmetry-invariant octahedral distortions remains a longstanding challenge in  these systems due to the weak x-ray scattering cross-section of oxygen atoms.  Consequently, complementary high-flux synchrotron x-ray diffraction and/or neutron diffraction techniques are often required to reliably resolve the weak structural distortions and their influence on the local octahedral framework ~\cite{Barnes_ACBS_06, Kennedy_AJC_12}. Temperature-dependent Raman spectroscopy provides an additional local probe of lattice dynamics and symmetry evolution, owing to the sensitivity of its vibrational modes to changes in the octahedral distortions, and has proven particularly useful for identifying subtle structural transitions in complex oxides ~\cite{Iliev_PRB_07, Castro_JRS_09, Manoun_JMS_12}. 

In this context, perovskite cobaltites have attracted significant attention because the multiple valence and spin states of cobalt ions give rise to a rich interplay among crystal structure, electronic configuration, and magnetism \cite{Viola_CM_02, Viola_CM_03, Yan_PRL_14, Morrow_JACS_13, Mabbs_book_73}. In octahedral coordination, Co$^{3+}$ ($3d^6$) can adopt  low-spin (LS, $t_{2g}^{6}e_{g}^{0}$, $S=0$), intermediate-spin (IS, $t_{2g}^{5}e_{g}^{1}$, $S=1$), or high-spin (HS, $t_{2g}^{4}e_{g}^{2}$, $S=2$) configurations, governed by the competition between crystal-field (CF) splitting and Hund's exchange energy. Consequently, subtle modifications of the local crystal structure can strongly influence the electronic and magnetic ground states. In particular, Co--O bond lengths and Co--O--B$'$ (or Co--O--Co) bond angles directly affect the CF strength and bandwidth, thereby tuning the spin-state energetics \cite{Chin_SR_17, Chen_JACS_14}. For example, a Co--O bond length exceeding 1.93 \AA\ has been reported to favor the HS state of Co$^{3+}$ in SrCo$_{0.5}$Ru$_{0.5}$O$_{3-\delta}$, whereas shorter bond lengths stabilize the LS state due to enhanced CF splitting \cite{Chen_JACS_14}. Similarly, tensile strain in LaCoO$_3$ elongates the Co--O bonds, reduces the CF splitting, and promotes the HS Co$^{3+}$, leading to the emergence of ferromagnetic order \cite{Fuchs_PRB_07, Li_NatComm_23}. Moreover, the partial occupation of antibonding $e_g$ orbitals in the IS state gives rise to pronounced Jahn--Teller distortions of the CoO$_6$ octahedra, lifting the orbital degeneracy and lowering the total energy of the system \cite{Kumar_JPCL_22, Radaelli_PRB_02, Maris_PRB_03}. As a result, temperature-induced spin-state transitions are frequently accompanied by significant structural modifications and symmetry changes, highlighting the delicate structure-property relationship in these compounds \cite{Yan_PRL_14, Vogt_PRL_00, Maris_PRB_03, Kozlenko_PRB_07}. In contrast, Co$^{2+}$ ions generally adopt the HS configuration (3$d^7$, $t_{2g}^{5}e_{g}^{2}$) in octahedral environments due to their weaker CF splitting, resulting in a reduced tendency toward cooperative Jahn--Teller distortions \cite{Mabbs_book_73, Viola_CM_03, Kumar_PRB1_20}. 

Within the perovskite cobaltites, the Sr$_{2-x}$La$_x$CoNbO$_6$ series provides an attractive platform for investigating the interplay among Co valence, $B$-site ordering, crystal structure, and magnetism, with Co being the sole magnetic ion in the system \cite{Kumar_PRB1_20, Kumar_PRB_22, Kumar_PRB2_20, Kumar_PRB_24, Kumar_JPCL_22}. Substitution of La$^{3+}$  at the  Sr$^{2+}$ site progressively reduces Co$^{3+}$ in Sr$_2$CoNbO$_6$ ($x=0$) to Co$^{2+}$ in SrLaCoNbO$_6$ ($x=1$), increasing the Co/Nb charge contrast and promoting rock-salt $B$-site ordering \cite{Kumar_PRB_22, Kumar_PRB1_20}.  Consequently, the series evolves from a nearly disordered tetragonal phase ($I4/m$) at low La content to a $B$-site ordered monoclinic phase ($P2_1/n$) for $x\geq0.6$, accompanied by a crossover from spin-glass behavior to long-range antiferromagnetic order \cite{Kumar_PRB1_20, Kumar_PRB2_20, Kumar_JPCL_22}. Among this series, SrLaCoNbO$_6$ is particularly intriguing because it hosts high-spin Co$^{2+}$ ions in a weak octahedral crystal field, resulting in a spin--orbit-coupled $^4T_1$ ground state with an effective pseudo-spin $\tilde{S}=1/2$ and a substantial unquenched orbital moment \cite{Kumar_PRB1_20, Kumar_PRB2_20, Mabbs_book_73, Lloret_ICA_08, Bos_PRB_04}. Previous studies have demonstrated that the ordered monoclinic $P2_1/n$ structure of SrLaCoNbO$_6$ remains stable down to 4 K and exhibits antiferromagnetic ordering below $T_\mathrm{N}\approx15$ K \cite{Bos_PRB_04,Kumar_PRB1_20,Kumar_PRB2_20}. However, our recent temperature-dependent Raman spectroscopy and Co $K$-edge EXAFS measurements revealed pronounced anomalies in the phonon frequencies, linewidths, and spectral weights near 60 K, together with additional modifications around 160--180 K, indicating a complex interplay between the spin and lattice degrees of freedom well above $T_{\mathrm N}$ \cite{Kumar_arXiv_26}. Neutron diffraction measurements further revealed significant geometric frustration arising from competing near-90$^\circ$ Co--O--Nb--O--Co superexchange pathways between Co$^{2+}$ moments \cite{Bos_PRB_04}. Despite extensive investigations of its low-temperature magnetic properties, the high-temperature structural behavior of SrLaCoNbO$_6$ remains unexplored. In particular, the evolution of the octahedral tilt network and the possible occurrence of structural phase transitions upon approaching the high-symmetry perovskite phase have not yet been investigated. 

In this work, we investigate the high-temperature structural changes of SrLaCoNbO$_6$ using temperature dependent Raman spectroscopy and high flux laboratory powder  XRD over the range 25--900$^\circ$C. We identify two distinct anomalies at $T_\mathrm{OD} = 300\pm 25~^\circ$C and $T_\mathrm{PT} = 750\pm 25~^\circ$C. Combined Raman and diffraction analyses reveal that the lower-temperature anomaly corresponds to an isostructural octahedral distortion within the monoclinic $P2_1/n$ phase, whereas the higher-temperature anomaly arises from a structural phase transition to the monoclinic $I2/m$ phase. Through group-theoretical analysis, Rietveld refinement, and quantitative evaluation of octahedral distortion parameters, we establish the structural origin of both anomalies and rule out several competing structural models. These results provide new insight into the temperature evolution of the octahedral framework in SrLaCoNbO$_6$ and demonstrate a robust laboratory-based approach for distinguishing local octahedral distortions from symmetry-breaking phase transitions in complex double perovskite oxides.

\section{\noindent ~Experimental}

Polycrystalline SrLaCoNbO$_6$ was synthesized by the conventional solid-state reaction method. Stoichiometric amounts of SrCO$_3$, La$_2$O$_3$, Nb$_2$O$_5$, and Co$_3$O$_4$ (Sigma-Aldrich/Alfa Aesar, purity $\geq$99.99\%) were thoroughly ground in an agate mortar, pelletized, calcined at 900~$^\circ$C for 2 h, and subsequently sintered at 1300~$^\circ$C for 48 h with intermediate regrindings \cite{Bos_PRB_04, Kumar_PRB1_20}. Field-emission scanning electron microscopy (FESEM) and energy-dispersive X-ray spectroscopy (EDS) elemental mapping were carried out using a TESCAN MAGNA LMU microscope equipped with an EDAX AMETEK Octane Elite Super EDS detector to investigate the surface morphology, elemental composition, and spatial distribution of the constituent elements. Prior to imaging, the sample was sputtered with a thin Au layer to ensure adequate electrical conductivity. High-resolution transmission electron microscopy (HRTEM) measurements were performed using an FEI Tecnai TF20 transmission electron microscope equipped with a field-emission gun (FEG) and operated at an accelerating voltage of 200 kV. Selected-area electron diffraction (SAED) patterns were acquired on the same instrument, and the corresponding diffraction data were analyzed using ImageJ software. Differential scanning calorimetry (DSC) measurements were performed using a NETZSCH STA 449 F1 Jupiter simultaneous thermal analyzer. The SrLaCoNbO$_6$ powder was measured between 100--900~$^{\circ}$C at a heating rate of 5~$^{\circ}$C/min under a high purity air atmosphere.

Temperature-dependent Raman spectroscopy measurements were performed over the range 21--800$^\circ$C ($\Delta$T = 50$^\circ$C) using the 514 nm line of an Ar$^+$ ion laser. The spectra were recorded in backscattering geometry using a Dilor XY triple-monochromator equipped with a Linkam TS1500 high-temperature optical furnace. A laser power of 5~mW with a spot size of $\sim$0.1~mm was employed to avoid possible local sample heating. At each temperature, the Raman spectra were calibrated using standard neon emission lines. The Raman spectra collected at each temperature were initially fitted using a superposition of Voigt profiles together with a spline background. The refined Voigt mixing parameter $\eta$ converged to unity across the entire measurement range, indicating that the phonon lineshapes are well described by pure Lorentzian functions.  This implies that the Gaussian contributions arising from the spectrometer resolution and from inhomogeneous broadening associated with disorder, defects, or strain are negligible. Subsequent fitting was therefore carried out using pure Lorentzian lineshapes. Two complementary fitting strategies were employed to extract the phonon parameters. In the first approach, the full spectrum was fitted simultaneously using a global spline background and a superposition of Lorentzian profiles. In the second approach, individual phonon modes were fitted locally over a restricted spectral window using a Lorentzian profile and a linear background, thereby minimizing parameter correlations arising from weak or spectrally overlapping neighboring features. The phonon frequencies $\omega_0$, linewidths $\Gamma$, and integrated intensities $I$ obtained from both procedures were consistent within the fitting uncertainties. Given that the local fitting procedure reduces inter-parameter correlations while yielding identical results, it was adopted for all subsequent analysis.

High-flux powder X-ray diffraction (XRD) measurements were carried out in Bragg–Brentano geometry over the range $\sim$25--900~$^\circ$C ($\Delta T = 25~^\circ$C) using a Rigaku SmartLab diffractometer equipped with Cu K$_\alpha$ and Mo K$_\alpha$ rotating-anode sources (45 kV, 200 mA) and an Anton Paar HTK 1200N high-temperature stage. Measurements using Mo K$_\alpha$ radiation were performed both in ambient atmospheric conditions and under high vacuum ($\sim$2--3 $\times$ 10$^{-6}$ mbar) to examine possible structural changes associated with oxygen-vacancy formation during high-temperature measurements. No detectable differences were observed between the diffraction patterns collected under the two conditions at any temperature. Therefore, the data recorded in the ambient conditions were used for the subsequent analyses for both radiation sources. Rietveld refinements were performed using the \textsc{FullProf} software package employing a pseudo-Voigt peak profile and linear interpolation between selected background points \cite{Carvajal_PB_93}. The room-temperature Cu K$_\alpha$ diffraction pattern was first refined to determine the instrumental parameters, which were subsequently fixed for all other temperatures and only the scale factor, lattice parameters, atomic coordinates, and overall isotropic displacement parameter ($B_{\rm iso}$) were allowed to vary during the refinements. Bond lengths, bond angles, and three-dimensional crystal-structure visualizations were obtained using the \textsc{VESTA} software package.

\begin{figure*}  
\centering
\includegraphics[width=1\linewidth]{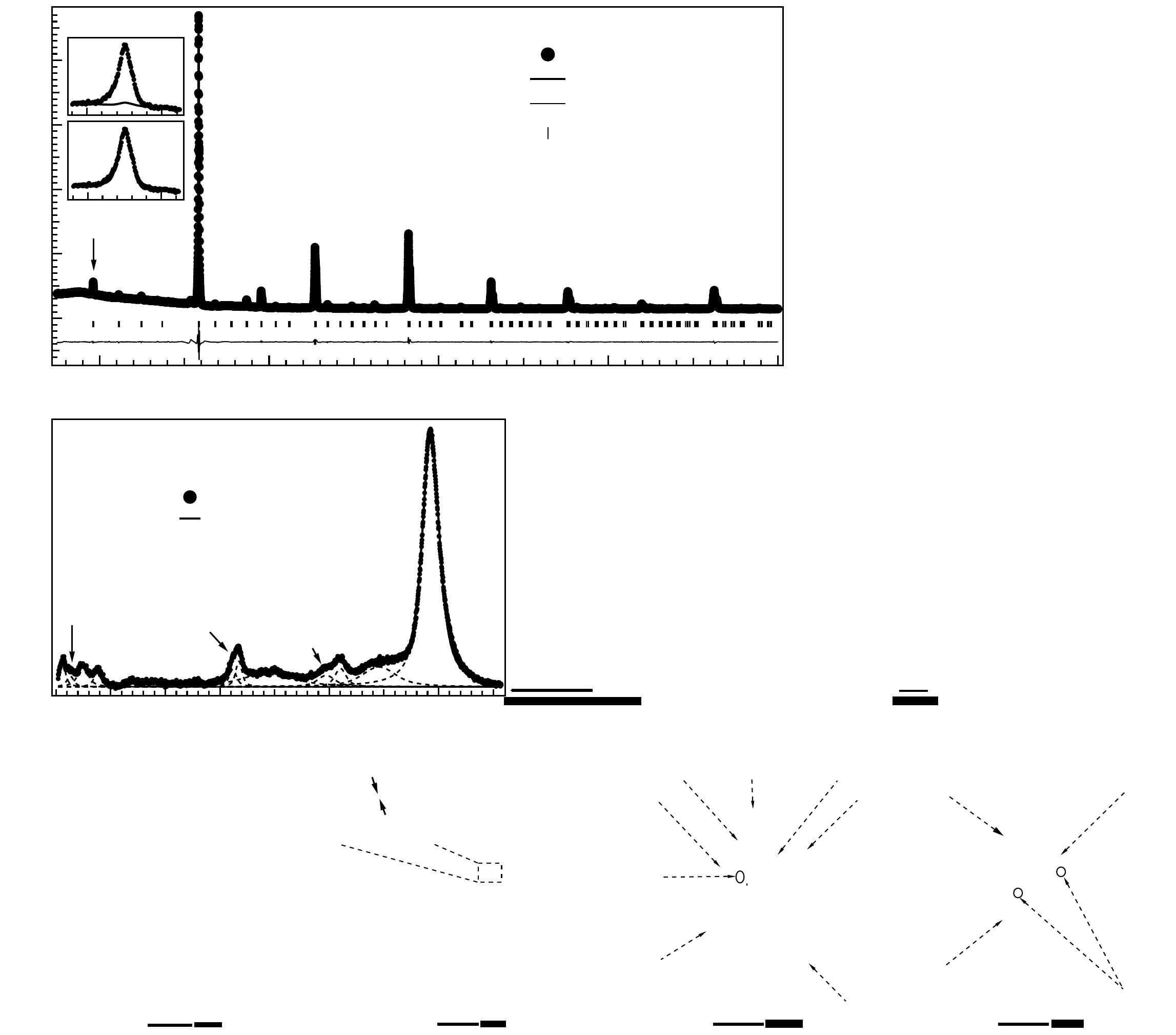}
\caption {(a) Rietveld refinement of the room-temperature XRD data of SrLaCoNbO$_6$ recorded using Cu K$_\alpha$ radiation. The red circles, solid black line, green vertical ticks, and solid blue line represent the experimental data points, fitted profile, Bragg reflection positions, and difference between the experimental and fitted curves, respectively. Insets (a1) and (a2) show the enlarged view of the (101) superlattice reflection fitted with B-site disordered and ordered structural models, respectively.  (b) Three-dimensional schematic representation of the room-temperature B-site ordered monoclinic crystal structure of SrLaCoNbO$_6$.  (c) Room-temperature Raman spectrum of the sample recorded using a 514 nm laser excitation source. The red circles, solid black curve, and dashed blue curves represent the observed spectrum, cumulative fitted spectrum, and individual mode fittings, respectively.  (d) FESEM image of the sample surface morphology.  (e--h) EDS elemental color maps of the same region, where the color intensity represents the relative characteristic X-ray counts for the indicated elemental lines. (i) HRTEM image of the sample.  (j) Higher magnification HRTEM image; the inset shows the inverse FFT image of the selected area.  (k, l) SAED patterns recorded from two different regions of the sample. The yellow circles indicate the diffraction spots corresponding to the (101) superlattice reflection. } 
\label{Fig1}
\end{figure*}

\begin{figure*}  
\centering
\includegraphics[width=1\linewidth]{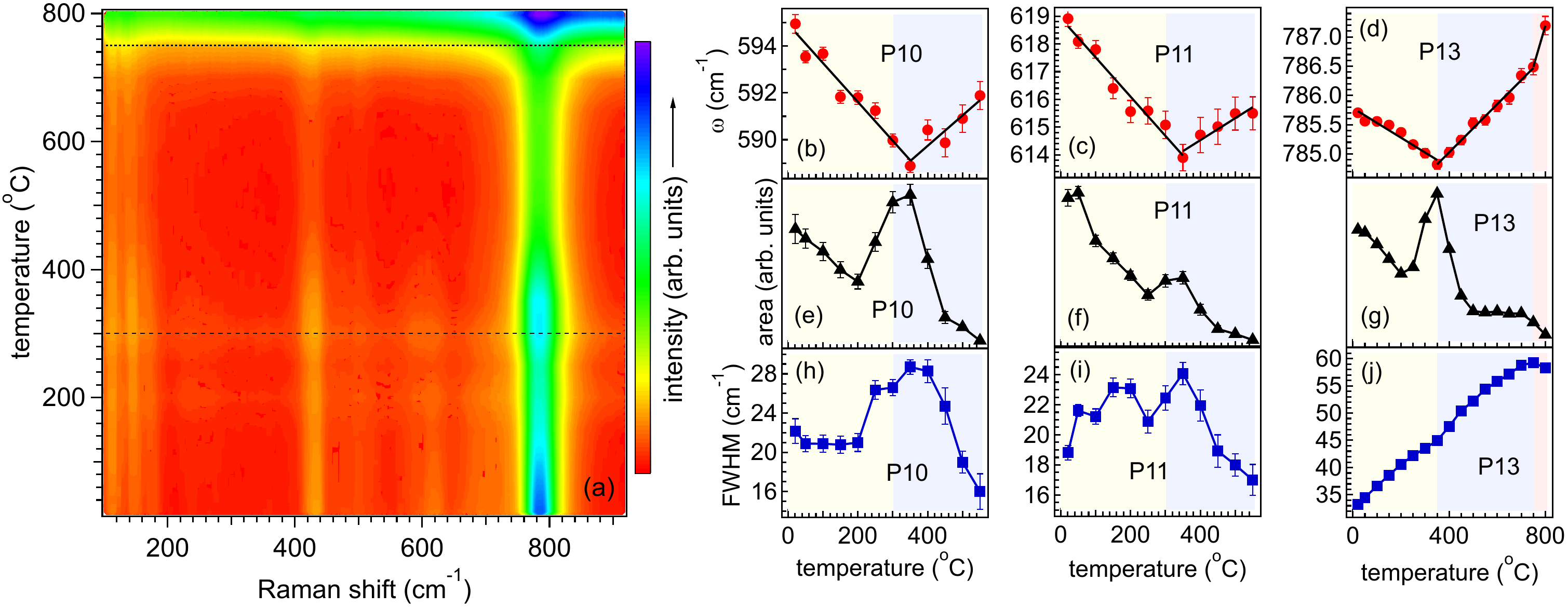}
\caption {(a) Contour plot of the temperature-dependent Raman spectra of SrLaCoNbO$_6$ recorded from 21~$^\circ$C to 800~$^\circ$C using a 514~nm excitation source. The horizontal dashed and dotted lines indicate anomalies in the spectral intensity around 300~$^\circ$C and 750~$^\circ$C, respectively. Temperature dependence of the (b--d) phonon frequency, (e--g) integrated area, and (h--j) linewidth of modes P10, P11, and P13, respectively. Solid black lines in (b--d) serve as guides to the eye. } 
\label{Fig2}
\end{figure*}

\section{\noindent ~Results }
\subsection{\noindent ~Room temperature crystal structure}

Figure~\ref{Fig1}(a) shows the Rietveld refinement of the room-temperature XRD pattern of SrLaCoNbO$_6$ recorded using Cu $K_{\alpha}$ radiation. The refinement confirms the formation of a monoclinic crystal structure with the $P2_1/n$ space group symmetry, consistent with previous reports~\cite{Kumar_PRB1_20,Bos_PRB_04}. The refined structural parameters at room temperature are listed in Table~\ref{T_XRD}. A distinct superlattice reflection corresponding to the (101) plane, marked by the arrow in Fig.~\ref{Fig1}(a), is clearly observed. This reflection arises from the long-range B-site ordering, i.e., the alternating arrangement of corner-sharing CoO$_6$ and NbO$_6$ octahedra within the crystal lattice, as schematically illustrated in Fig.~\ref{Fig1}(b). To further verify this, enlarged views of the (101) superlattice reflection fitted using fully disordered and fully ordered structural models are presented in the insets  (a1) and (a2) of Fig.~\ref{Fig1}(a), respectively, while keeping all other refinement parameters unchanged. The excellent agreement between the observed and calculated intensities of the superlattice reflection shown in inset (a2) confirms the highly ordered arrangement of Co and Nb ions in the structure at room temperature.

The room-temperature Raman spectrum of the sample recorded using 514~nm laser line for excitation is displayed in Fig.~\ref{Fig1}(c).  Group-theoretical analysis provides  24 Raman-active (12~$A_g$ + 12~$B_g$) and 36 infrared-active modes for the monoclinic $P2_1/n$ structure with $C_{2h}$ point symmetry (see Table~S1 of \cite{SI}), with the Raman-active vibrations described by the irreducible representation $\Gamma_{\rm Raman}=\nu_1(A_g+B_g)+\nu_2(2A_g+2B_g)+\nu_5(3A_g+3B_g)+T(3A_g+3B_g)+L(3A_g+3B_g)$~\cite{Singh_PRR1_20,Singh_PRM_20}. Here, $\nu_1$, $\nu_2$, and $\nu_5$ correspond to the internal vibrations of the (Co/Nb)O$_6$ octahedra, while $T$ and $L$ represent translational and librational modes, respectively. The observed Raman peaks, labeled P1--P13 in Fig.~\ref{Fig1}(c), can be classified into three frequency regions. The high-frequency modes P10--P13 ($\sim$550--850~cm$^{-1}$) originate from internal (Co/Nb)--O stretching vibrations within the (Co/Nb)O$_6$ octahedra. The intermediate-frequency modes P5--P9 (200--550~cm$^{-1}$) are primarily associated with oxygen bending vibrations, corresponding to octahedral rotations (i.e., modulation of the Co/Nb--O--Co/Nb bond angles).  The low-frequency modes P1--P4, (below $\sim$200~cm$^{-1}$) arise from translational motion of the A-site cations and octahedral lattice vibrations~\cite{Andrews_DT_15,Iliev_PRB_07,Ayala_JAP_07}.

To examine the surface morphology and elemental distribution, FESEM imaging and EDS elemental mapping were performed. The FESEM image [Fig.~\ref{Fig1}(d)] reveals micron-sized grains, as typically expected for samples synthesized via the solid-state reaction route. The corresponding EDS elemental maps [Figs.~\ref{Fig1}(e--h)] confirm the homogeneous distribution of Sr, La, Co, and Nb throughout the examined region. Notably, the common dark regions observed in all elemental maps arise from morphology-related attenuation and shadowing of the X-ray signal, rather than from depletion of the constituent elements. To further investigate the microstructural properties, HRTEM measurements were performed. The HRTEM image [Fig.~\ref{Fig1}(i)] reveals well-defined crystalline grains, while the higher-magnification image [Fig.~\ref{Fig1}(j)] exhibits clear lattice fringes, confirming the excellent crystallinity of the sample. The inverse fast Fourier transform (IFFT) image of the region enclosed by the red dashed rectangle is shown in the inset of Fig.~\ref{Fig1}(j) for clarity. The average interplanar spacing, determined from several lattice fringes, corresponds to the (020) family of planes. The room-temperature selected-area electron diffraction (SAED) pattern [Fig.~\ref{Fig1}(k)] exhibits distinct diffraction rings indexed to different crystallographic planes. In addition, weak diffraction spots corresponding to the superlattice (101) reflection are visible and are highlighted by the yellow circles. To further verify the presence of this superlattice reflection, SAED measurements were performed at another region of the sample [Fig.~\ref{Fig1}(l)]. The appearance of well-defined diffraction spots associated with the (101) planes, highlighted by the yellow circles, confirms the B-site-ordered crystal structure of SrLaCoNbO$_6$ at room temperature, in agreement with the XRD analysis.

\begin{figure*}  
\centering
\includegraphics[width=0.8\linewidth]{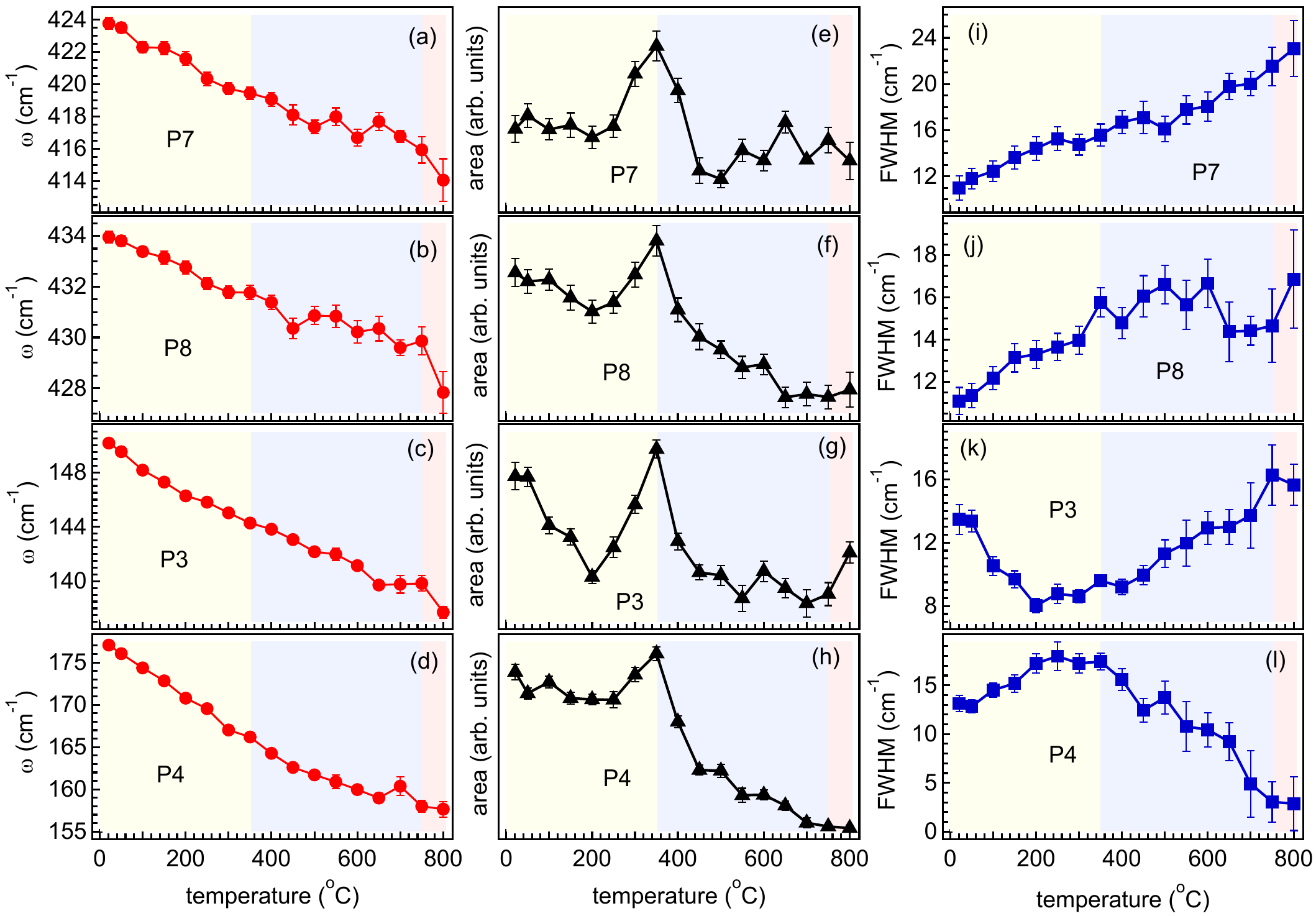}
\caption {Temperature dependence of the (a--d) phonon frequencies, (e--h) integrated intensities (area), and (i--l) linewidths of modes P7, P8, P3, and P4, respectively. } 
\label{Fig3}
\end{figure*}

\subsection{\noindent ~High temperature crystal structure}
\subsubsection{\noindent ~Raman spectroscopy}

A contour map of the high-temperature Raman spectra recorded from 21~$^\circ$C to 800~$^\circ$C using a 514~nm excitation wavelength is shown in Fig.~\ref{Fig2}(a) [see Fig.~\ref{Fig13}(a) of the Appendix for representative line profiles]. Interestingly, a distinct anomaly in the intensities of all prominent Raman modes is observed at $T = 300\pm 50~^\circ$C, as indicated by the horizontal dashed line in Fig.~\ref{Fig2}(a). Near this temperature, the mode intensities initially increase upon heating and subsequently decrease with further increase in temperature (see Fig.~S1 of \cite{SI} for clarity). This behavior contrasts with the conventional thermal response of Raman-active phonons, in which anharmonic phonon scattering and increasing thermal disorder typically lead to a continuous reduction in mode intensity with increasing temperature. A second, more pronounced increase in the spectral intensity is observed at $750\pm 50~^\circ$C, as marked by the dotted line in Fig.~\ref{Fig2}(a). To investigate the origin of these anomalies, the Raman spectra at each temperature were fitted (see Fig.~S2 of \cite{SI}), and the temperature dependence of the phonon frequency, integrated intensity (area), and linewidth of the high-energy modes P10, P11, and P13 is presented in Figs.~\ref{Fig2}(b--d), \ref{Fig2}(e--g), and \ref{Fig2}(h--j), respectively. All three modes exhibit clear anomalies in their phonon self-energy parameters [frequency ($\omega$) and linewidth ($\Gamma$)] near 350~$^\circ$C, indicating a structural transformation. Modes P10 and P11 exhibit a redshift accompanied by weak, non-monotonic linewidth broadening up to $\sim$350~$^\circ$C. Above this temperature, both modes display a blueshift together with pronounced linewidth narrowing [see Figs.~\ref{Fig2}(b,c) and \ref{Fig2}(h,i)]. Furthermore, P10 and P11 remain discernible up to $\sim$700~$^\circ$C but disappear at higher temperatures [see Fig.~\ref{Fig13}(d) of the Appendix], suggesting that the high-temperature anomaly is associated with an increase in the crystal symmetry. Owing to the progressive broadening and weakening of these modes at elevated temperatures, a reliable fitting was possible only up to 550~$^\circ$C [Fig.~\ref{Fig13}(c)].

Moreover, the integrated intensities of modes P10 and P11 [Figs.~\ref{Fig2}(e,f)] initially decrease with increasing temperature, exhibit an abrupt enhancement near 350~$^\circ$C, and subsequently decrease upon further heating  [see also Figs.~\ref{Fig13}(e--g) of the Appendix]. Furthermore, the intensity of mode P10 increases relative to that of P11 with increasing temperature, and P10 becomes the dominant feature near 300~$^\circ$C before rapidly weakening  for $T \gtrsim$ 350~$^\circ$C [see Figs.~\ref{Fig13}(b, c) and \ref{Fig13}(e--g) of the Appendix]. The corresponding intensity ratio, $I_{\rm P10}/I_{\rm P11}$ [inset of Fig.~\ref{Fig13}(f)], exhibits an abrupt change near 300~$^\circ$C, further supporting a distinct structural rearrangement at this temperature. The weak mode P12 becomes nearly unobservable above $\sim$300~$^\circ$C and exhibits a monotonic blueshift accompanied by linewidth narrowing between 21 and 250~$^\circ$C [Figs.~S3(a--d) of \cite{SI}]. The most intense mode, P13, whose phonon parameters can be reliably extracted over the entire temperature range, also displays a crossover from a redshift below 350~$^\circ$C to a blueshift above this temperature [Fig.~\ref{Fig2}(d)], accompanied by a pronounced enhancement of its integrated intensity [Fig.~\ref{Fig2}(g); see also Figs.~S3(g--i) of \cite{SI}], analogous to the behavior of P10 and P11. In contrast to its phonon frequency and integrated intensity, the linewidth of P13 increases nearly monotonically with temperature and exhibits only a weak anomaly near 350~$^\circ$C [Fig.~\ref{Fig2}(j)].

Since these high-energy modes originate primarily from the stretching vibrations of oxygen atoms within the (Co/Nb)O$_6$ octahedra, the observed anomalies in their frequencies, linewidths, and intensities strongly indicate a modification of the octahedral framework involving changes in the tilting and/or rotational distortions around 350~$^\circ$C. Although the Raman results provide compelling evidence for a structural rearrangement, they do not by themselves prove a change in crystallographic symmetry. Nevertheless, the enhancement of the Raman intensity together with the concomitant renormalization of the phonon self-energy parameters near 300-350~$^\circ$C is consistent with the evolution toward a less distorted octahedral configuration at higher temperatures.

An additional anomaly in the parameters of mode P13 is also observed around $\sim$750~$^\circ$C [see Figs.~\ref{Fig2}(d,g,j)]. This, together with the disappearance of modes P10 and P11, points to the possibility of a second structural transition near this temperature. Nevertheless, measurements at higher temperatures are required to establish its nature unambiguously. It is noteworthy that, although the overall spectral intensity of mode P13 increases markedly above $\sim$700~$^\circ$C [see Fig.~\ref{Fig2}(a) and Fig.~S3(j) of \cite{SI}], the actual area under the P13 phonon mode decreases [Fig.~\ref{Fig2}(g)]. This apparent inconsistency arises from a substantial increase in the underlying spectral background at elevated temperatures, which contributes significantly to the total scattered intensity. After subtraction of a linear background contribution [see Fig.~S2(d1--d6)], the integrated intensity associated with the P13 phonon is found to decrease above $\sim$700~$^\circ$C. This behavior is in contrast to that observed near 300~$^\circ$C, where the enhancement in spectral intensity originates from a genuine increase in the phonon's intrinsic intensity.

The temperature dependence of $\omega$, $I$, and $\Gamma$ for the intermediate-energy bending modes (P7 and P8) [Fig.~S3(e)] and the low-energy translational modes (P3 and P4) [Fig.~S3(f)] is presented in Figs.~\ref{Fig3}(a--d), \ref{Fig3}(e--h), and \ref{Fig3}(i--l), respectively. In contrast to the high-energy stretching modes, these modes do not exhibit any anomaly in their phonon frequencies near 350~$^\circ$C. Given that bending and translational vibrations are considerably less sensitive to local distortions of the (Co/Nb)O$_6$ octahedra than stretching vibrations, this observation suggests that the anomaly near 350~$^\circ$C is primarily associated with a modification of the octahedral environment. Despite the absence of a pronounced frequency anomaly, all four modes exhibit a significant enhancement in their integrated intensities around 350~$^\circ$C [see Figs.~\ref{Fig3}(e--h)], consistent with the behavior of the high-energy stretching modes. In addition, subtle changes in the linewidth are also detected near this temperature, particularly for the translational modes P3 and P4 [Fig.~\ref{Fig3}(i--l)]. Collectively, these results provide strong spectroscopic evidence for a reorganization of the octahedral distortion pattern near $350\pm 50~^\circ$C and suggest the possibility of a higher-temperature structural phase transition near $750\pm 50~^\circ$C.

Factor-group analysis predicts only four Raman-active modes for the cubic $Fm\bar{3}m$ structure with the $a^-a^-a^-$ tilt system, described by the irreducible representation $\Gamma_g(Fm\bar{3}m)=\nu_1(A_{1g})+\nu_2(E_g)+\nu_5(F_{2g})+T(F_{2g})$~\cite{Ayala_JAP_07,Andrews_DT_15}. However, at least seven Raman modes (P1, P3, P4, P7, P8, P9, and P13) remain clearly visible above 700~$^\circ$C [see Fig.~\ref{Fig13}(a)], thereby ruling out the transformation into a cubic $Fm\bar{3}m$ phase at high temperature. Likewise, the rhombohedral $R\bar{3}c$ and tetragonal $I4/m$ structures, which are commonly reported as high-temperature phases in double perovskites, are characterized by the irreducible representations $\Gamma_g(R\bar{3}c)=\nu_1(A_{1g})+\nu_2(E_g)+2\nu_5(A_g+E_g)+2L(A_g+E_g)+2T(A_g+E_g)$ and $\Gamma_g(I4/m)=\nu_1(A_{1g})+2\nu_2(A_g+B_g)+2\nu_5(B_g+E_g)+2L(A_g+E_g)+2T(B_g+E_g)$, respectively~\cite{Ayala_JAP_07,Andrews_DT_15,Castro_JRS_09}. Notably, both structures permit only two translational Raman modes. In contrast, the three low-energy translational modes P1, P3, and P4 remain clearly observable above $\sim$700~$^\circ$C, indicating that the high-temperature phase cannot be described by either $R\bar{3}c$ or $I4/m$ symmetry. Instead, the Raman results suggest a transition from the $P2_1/n$ phase to the monoclinic $I2/m$ structure at $750\pm50$~$^\circ$C. This transition temperature falls within the broad range reported for $P2_1/n\rightarrow I2/m$ transformations in related ordered double perovskites, namely Sr$_2$ScSbO$_6$ ($\sim$130~$^\circ$C)~\cite{Faik_JSSC_12}, Sr$_2$GdRuO$_6$ (300~$^\circ$C)~\cite{Triana_MRB_11}, Sr$_2$InTaO$_6$ (605~$^\circ$C)~\cite{Zhou_PCM_13}, Sr$_2$YTaO$_6$ (850~$^\circ$C)~\cite{Zhou_JSSC_10}, and Ca$_2$ScSbO$_6$ ($\sim$1170~$^\circ$C)~\cite{Faik_JSSC_12}.
The corresponding irreducible representation, $\Gamma_g(I2/m)=\nu_1(A_g)+\nu_2(2A_g)+3\nu_5(A_g+2B_g)+3L(A_g+2B_g)+3T(2A_g+B_g)$, allows three translational modes, in agreement with the experimental Raman spectra.

\subsubsection{\noindent ~X-ray diffraction}

\begin{figure*}  
\centering
\includegraphics[width=1\linewidth]{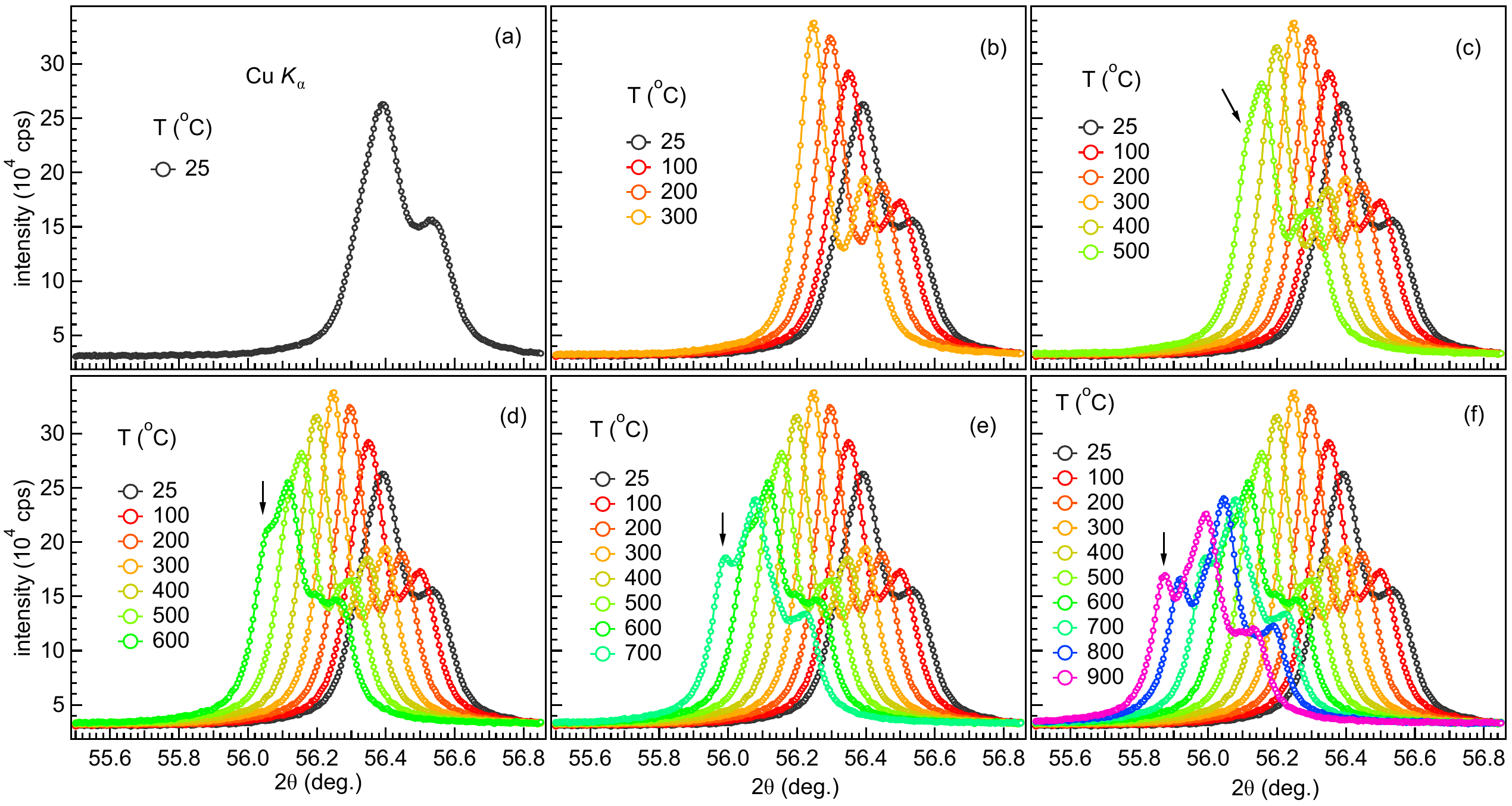}
\caption {(a--f) Selected XRD reflections measured using Cu $K_{\alpha}$ radiation, displayed with the successive addition of diffraction patterns acquired at increasing temperatures. The black arrows mark the additional reflections that emerge above $\sim$400~$^\circ$C.  } 
\label{Fig4}
\end{figure*}

\begin{table*}[t]
\centering
\caption{Structural parameters of SrLaCoNbO$_6$ obtained from Rietveld refinement of Cu $K_{\alpha}$ XRD data collected at room temperature ($P2_1/n$) and 900~$^\circ$C ($I2/m$). Numbers in parentheses are the estimated standard deviations in the last significant digits.$^\#$}
\label{T_XRD}

\renewcommand{\arraystretch}{1.0}
\begin{tabular*}{\textwidth}{@{\extracolsep{\fill}}lcc@{}}
\hline\hline
Parameter & RT ($P2_1/n$) & 900~$^\circ$C ($I2/m$)\\
\hline
$a$ (\AA) & 5.6428(2) & 5.7103(1) \\
$b$ (\AA) & 5.6524(1) & 5.6934(1) \\
$c$ (\AA) & 7.9767(1) & 8.0536(1) \\
$\beta$ ($^\circ$) & 89.89(1) & 89.82(1) \\
$V$ (\AA$^3$) & 254.42(3) & 261.82(4) \\[1mm]

Sr/La $(x,y,z)$ &
0.5066(4), 0.4742(3), 0.2504(6) &
0.5045(8), $\frac12$, 0.2509(7) \\

O1 $(x,y,z)$ &
0.7041(12), 0.6894(9), 0.0281(7) &
0.7078(11), 0.7240(13), 0.0393(9) \\

O2 $(x,y,z)$ &
0.7770(8), 0.2194(11), 0.0397(9) &
0.4890(10), 0, 0.2530(13) \\

O3 $(x,y,z)$ &
0.4404(13), 0.0138(10), 0.2495(8) &
--- \\[1mm]

$R_{\mathrm{Bragg}}$ (\%) & 2.28 & 2.81 \\
$R_F$ (\%) & 3.32 & 3.76 \\
\hline\hline
\end{tabular*}

\vspace{1mm}

\vspace{1mm}
\raggedright
\footnotesize
$^\#$RT ($P2_1/n$): La/Sr, O1, O2, O3 at 4$e$ $(x,y,z)$, Co at $2c$ $(\frac12,0,0)$ and Nb at $2d$ $(0,\frac12,0)$.  900~$^\circ$C ($I2/m$): La/Sr at 4$e$ $(x,\frac12,z)$, O1 at 8$f$ $(x,y,z)$, O2 at 4$e$ $(x,0,z)$, Co at $2c$ $(\frac12,0,0)$ and Nb at $2d$ $(0,\frac12,0)$.

\end{table*}

To elucidate the origin of the anomalous renormalization of the Raman phonon self-energy parameters, high-flux powder XRD measurements were performed on SrLaCoNbO$_6$ from $\sim$25 to 900 $^\circ$C using both Cu and Mo K$_\alpha$ radiations [see Figs.~S4(a, b) of Ref.~\cite{SI}]. Interestingly, analogous to the Raman modes, the intensities of most of the XRD peaks increase up to around 300~$^\circ$C and then decrease with further heating [see Figs.~S4(c, d)]. This suggests an enhancement in the crystallinity and/or structural symmetry of the compound with increasing temperature up to $\sim$300 $^\circ$C. In contrast, several new reflections emerge above $\sim$400 $^\circ$C, as indicated by the arrows in Figs.~\ref{Fig4}(a--f) (see Figs.~S5--S8 for other reflections). More importantly, none of the peaks corresponding to the room-temperature $P2_1/n$ symmetry disappear around this temperature. This observation is intriguing because crystal symmetry is generally expected to increase with temperature, which would typically result in the suppression of certain reflections at higher temperatures \cite{Mullens_CM_24, Martin_JMC_05}. Therefore, the appearance of additional reflections while retaining all peaks associated with the $P2_1/n$ symmetry suggests either a lowering of the crystal symmetry or a substantial lattice rearrangement within the same structure type. Since crystal symmetry of perovskite oxides generally increases with temperature \cite{Faik_JSSC_12, Martin_JMC_05, Gateshki_JPCM_03}, an increase in octahedral distortion, tilting, and/or rotation within the $P2_1/n$ structure appears to provide the most plausible explanation for the observed behavior.

\begin{figure}  
\centering
\includegraphics[width=0.85\linewidth]{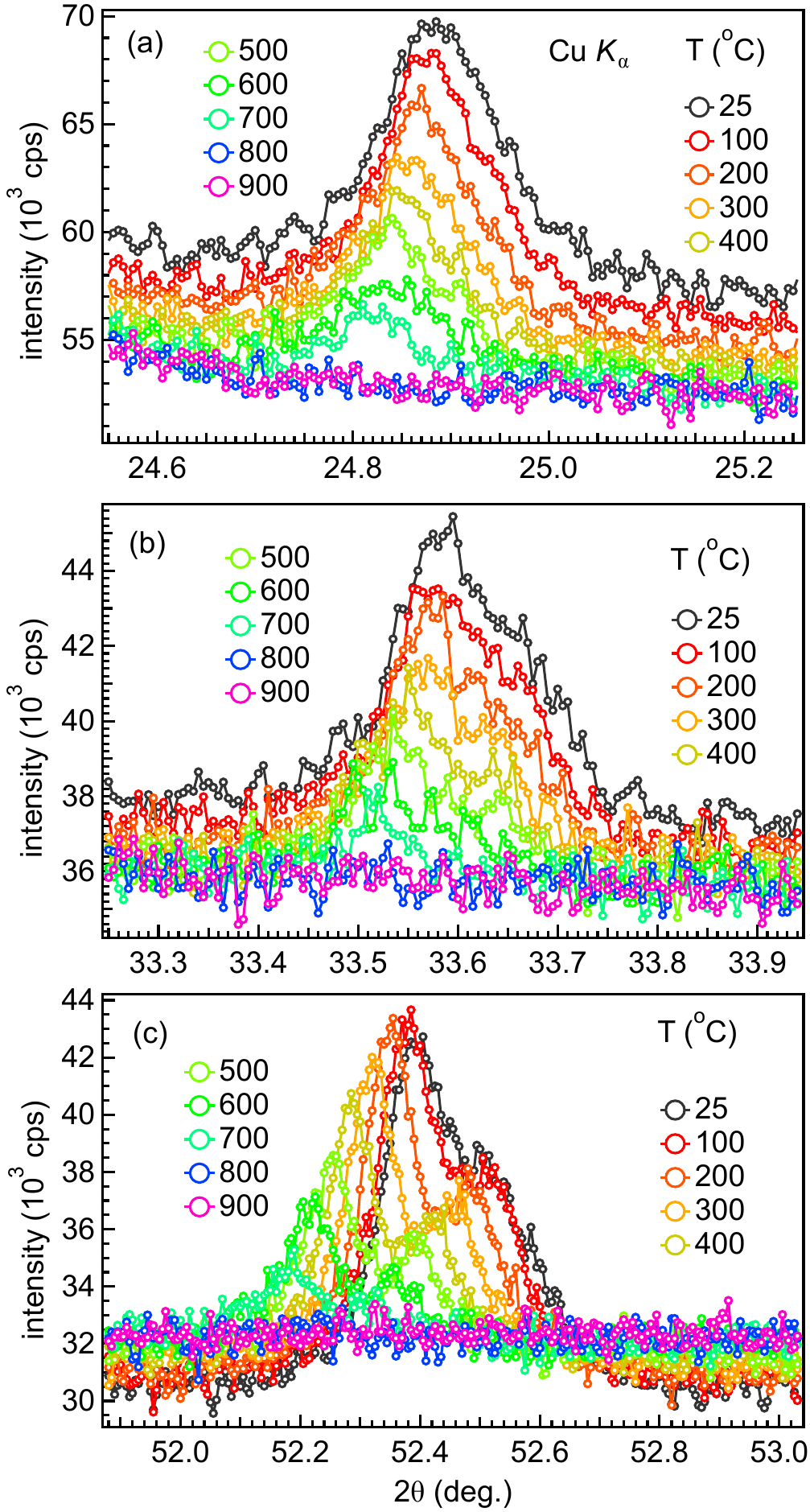}
\caption {(a--c) Enlarged views of selected weak Bragg peaks from the XRD patterns of SrLaCoNbO$_6$ measured at different temperatures using the Cu $K_{\alpha}$ radiation.} 
\label{Fig5}
\end{figure}

\begin{figure}  
\centering
\includegraphics[width=1\linewidth]{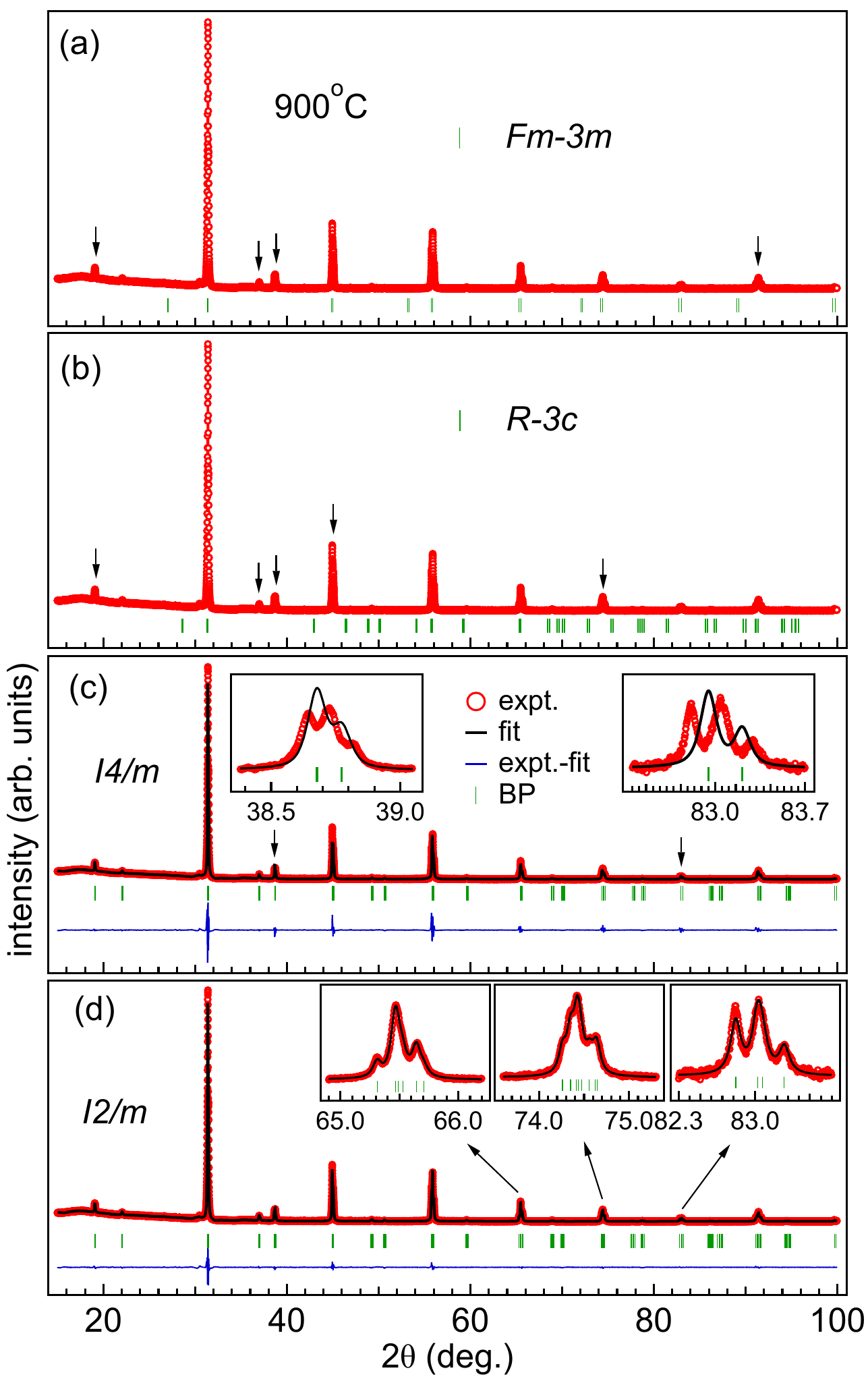}
\caption {XRD patterns of SrLaCoNbO$_6$ recorded at 900~$^\circ$C using Cu $K_{\alpha}$ radiation. Vertical green bars indicate the Bragg peak positions corresponding to the (a) $Fm\bar{3}m$, (b) $R\bar{3}c$, (c) $I4/m$, and (d) $I2/m$ structural models. The vertical arrows in panels (a) and (b) indicate experimentally observed reflections for which no Bragg positions are allowed by the corresponding space-group symmetries. Panels (c) and (d) show the Rietveld refinements using the $I4/m$ and $I2/m$ structural models, respectively. The insets show enlarged views of the reflections marked by the arrows.} 
\label{Fig6}
\end{figure}

Moreover, several XRD reflections disappear above $\sim$700$^\circ$C, as shown in Figs.~\ref{Fig5}(a--c) [see the red arrows in Fig.~S5(a,b) and Fig.~S9]. The systematic extinction of these reflections above a certain temperature suggests an increase in the crystal symmetry of the sample within this temperature range. To identify the high-temperature crystal structure, Rietveld refinements of the XRD pattern collected at 900 $^\circ$C were performed using several candidate structural models, as shown in Figs.~\ref{Fig6}(a--d). The vertical green ticks in Figs.~\ref{Fig6}(a) and \ref{Fig6}(b) denote the calculated Bragg peak positions for the cubic $Fm\bar{3}m$ (No.~225) and rhombohedral $R\bar{3}c$ (No.~167) structural models, respectively. It is evident that several experimentally observed reflections, marked by the vertical arrows, are forbidden by the reflection conditions of these space groups. This discrepancy rules out both the $Fm\bar{3}m$ and $R\bar{3}c$ models as possible high-temperature crystal structures, indicating that the high-temperature phase possesses a lower symmetry than either model. Figure~\ref{Fig6}(c) shows the Rietveld refinement using the tetragonal $I4/m$ (No.~87) model. Although this model reproduces the overall diffraction pattern reasonably well, closer inspection of selected reflections, particularly at higher $2\theta$ values, reveals noticeable discrepancies between the observed and calculated profiles, as highlighted in the insets. In particular, the additional splitting observed for several high-$2\theta$ reflections indicates that the high-temperature phase still retains a monoclinic distortion.

\begin{figure*} 
\centering
\includegraphics[width=0.80\linewidth]{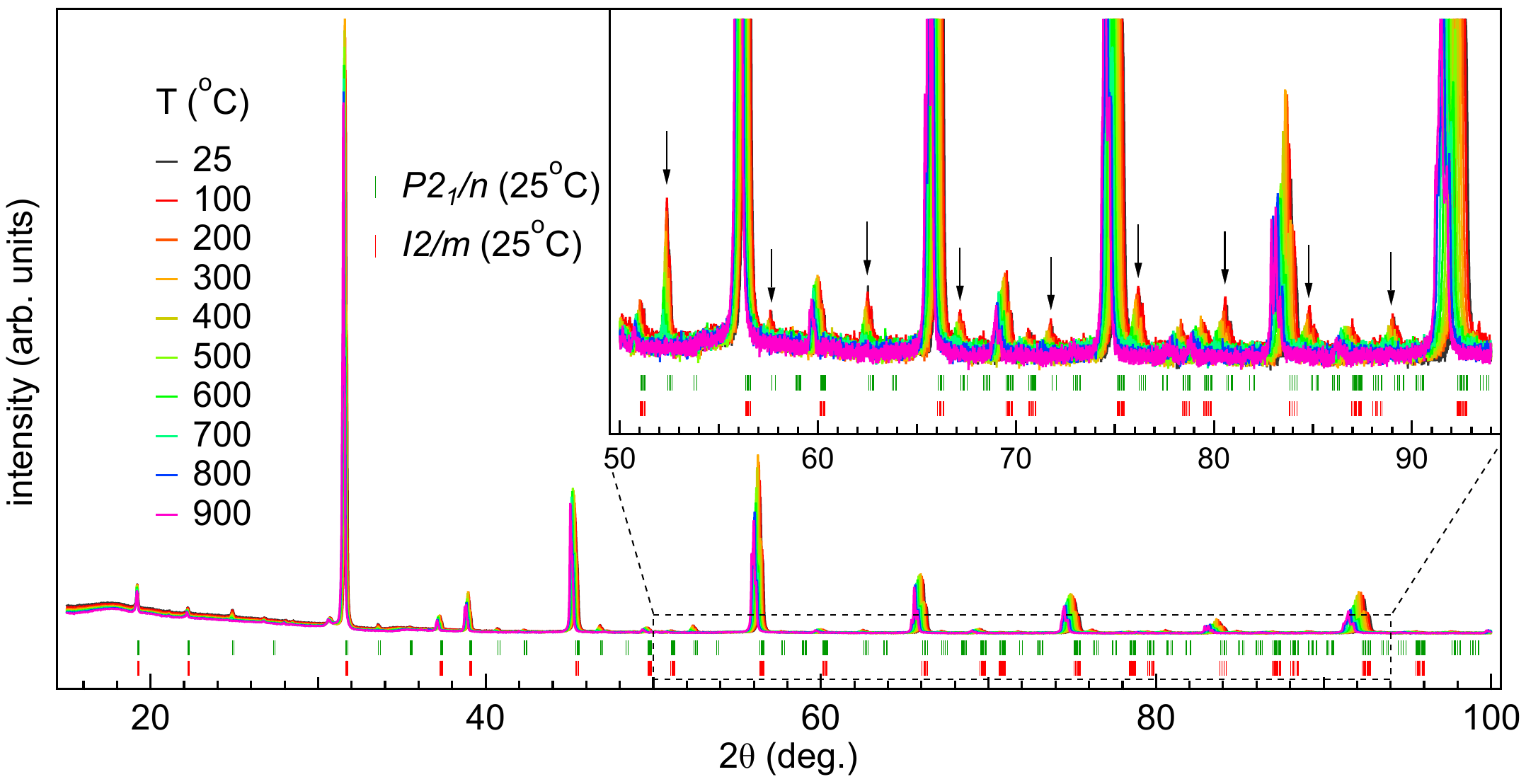} 
\caption {XRD patterns of SrLaCoNbO$_6$ collected at various temperatures using the Cu $K_{\alpha}$ radiation. The Bragg peak positions corresponding to the $P2_1/n$ and $I2/m$ phases, both calculated using the structural parameters obtained at 25~$^\circ$C, are shown for direct comparison. The inset displays an enlarged view of the high-$2\theta$ region, where the vertical arrows indicate reflections that vanish upon increasing temperature.}
\label{Fig7}
\end{figure*}

\begin{figure*}  
\centering
\includegraphics[width=0.85\linewidth]{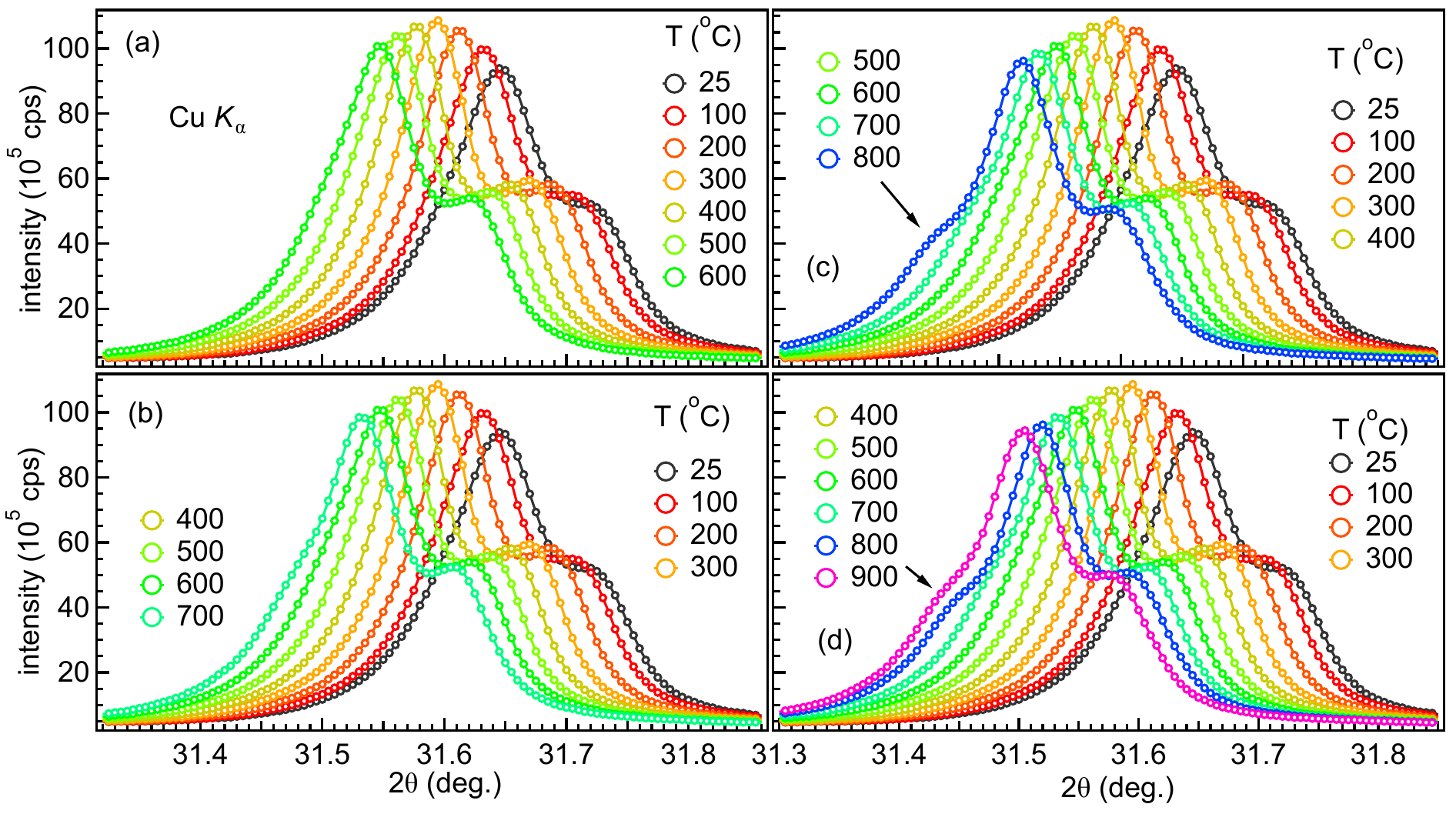}
\caption {(a--d) Selected XRD reflections measured using the Cu $K_{\alpha}$ radiation, displayed with the successive addition of diffraction patterns recorded at increasing temperatures starting from 600~$^\circ$C [in (a)]. The black arrows indicate the additional reflections appearing above $\sim$700~$^\circ$C.} 
\label{Fig8}
\end{figure*}

To further investigate this, several candidate monoclinic space groups were examined. A detailed analysis of the diffraction data reveals that the high-temperature ($T \gtrsim 750^\circ$C) patterns can be satisfactorily refined using the monoclinic $I2/m$ (No.~12) space group, as shown in Fig.~\ref{Fig6}(d). The insets display enlarged views of selected reflections in the high-$2\theta$ region, demonstrating the excellent agreement between the observed and calculated diffraction patterns. The corresponding refinement parameters are given in Table \ref{T_XRD}. This assignment is further supported by a direct comparison of the Bragg positions corresponding to the $P2_1/n$ and $I2/m$ phases in Fig.~\ref{Fig7}. The enlarged view of the high-$2\theta$ region clearly demonstrates that no Bragg position corresponding to the $I2/m$ phase is predicted for the reflections that disappear at high temperature, as indicated by the vertical arrows. In contrast, calculated Bragg positions are present for all reflections that persist in the high-temperature phase. For a direct comparison, the Bragg positions for both structural models were calculated using the refined structural parameters obtained at 25 $^\circ$C. More importantly, all reflections that become extinct at high temperature satisfy the condition $h+k+l=2n+1$ (odd), which is consistent with the reflection conditions associated with the $P2_1/n \rightarrow I2/m$ phase transformation. Therefore, we identify this temperature as the phase-transition temperature, $T_{\rm PT} = 750\pm 25~^\circ$C. Interestingly, the temperature-induced phase transition from the $P2_1/n$ to the $I2/m$ structure is also accompanied by the emergence of several additional Bragg reflections, as shown in Fig.~\ref{Fig8} (see also Figs.~S10 and S11), which suggests that the transition to the higher-symmetry phase is accompanied by significant modifications of the BO$_6$ octahedral network.

\begin{figure*}  
\centering
\includegraphics[width=0.75\linewidth]{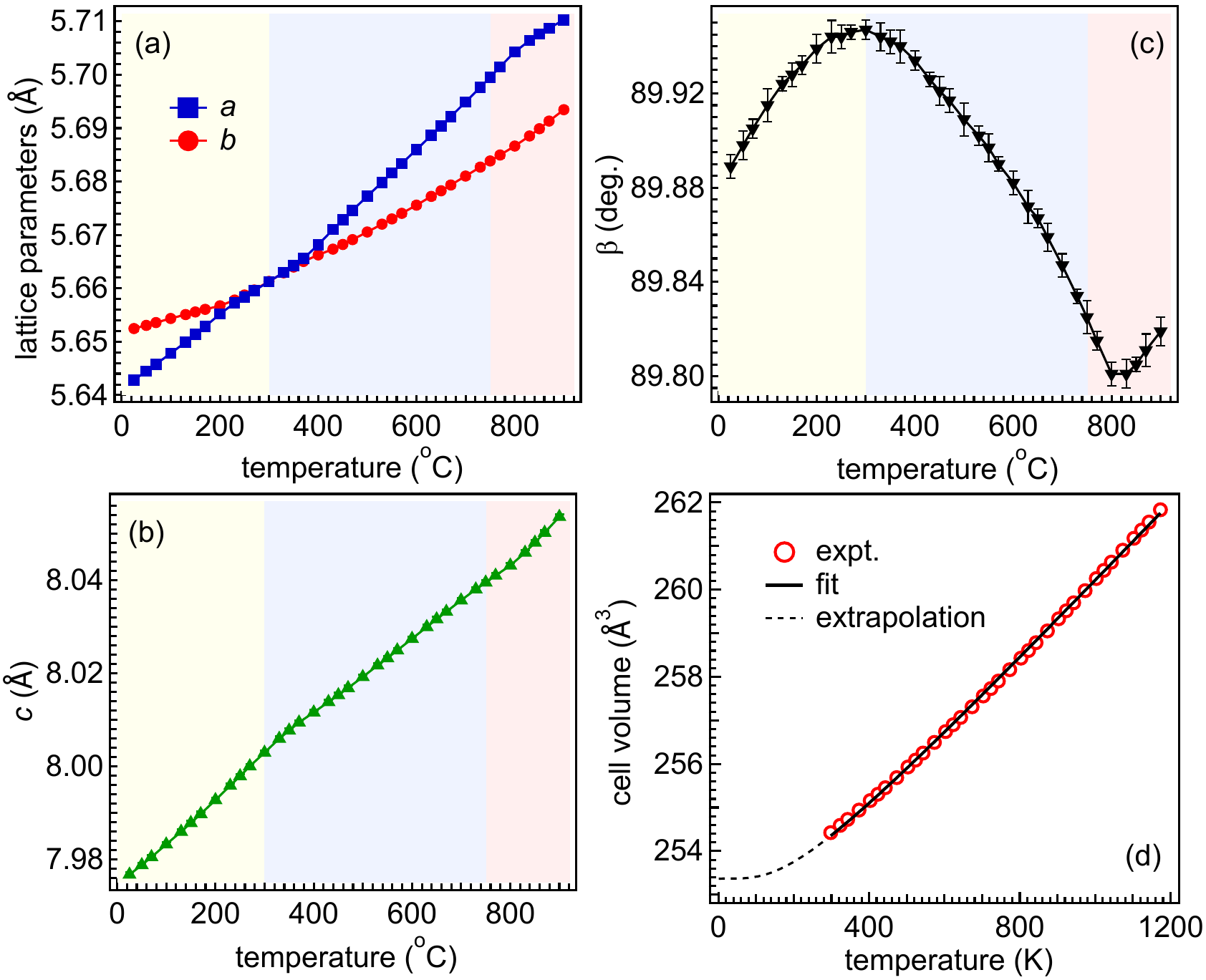}
\caption {Temperature dependence of the lattice parameters: (a) $a$ and $b$, (b) $c$, (c) the monoclinic angle $\beta$, and (d) the unit-cell volume of SrLaCoNbO$_6$, obtained from Rietveld refinements of the XRD data using the $P2_1/n$ space group. The error bars in panels (a), (b), and (d) are smaller than the symbol size. The solid black curve in panel (d) represents the best fit to the experimental data using Eqs.~\ref{Gur1} and \ref{Gur2}, while the dashed curve shows the extrapolation of the fit to 0~K.} 
\label{Fig9}
\end{figure*}

\begin{figure*}  
\centering
\includegraphics[width=1\linewidth]{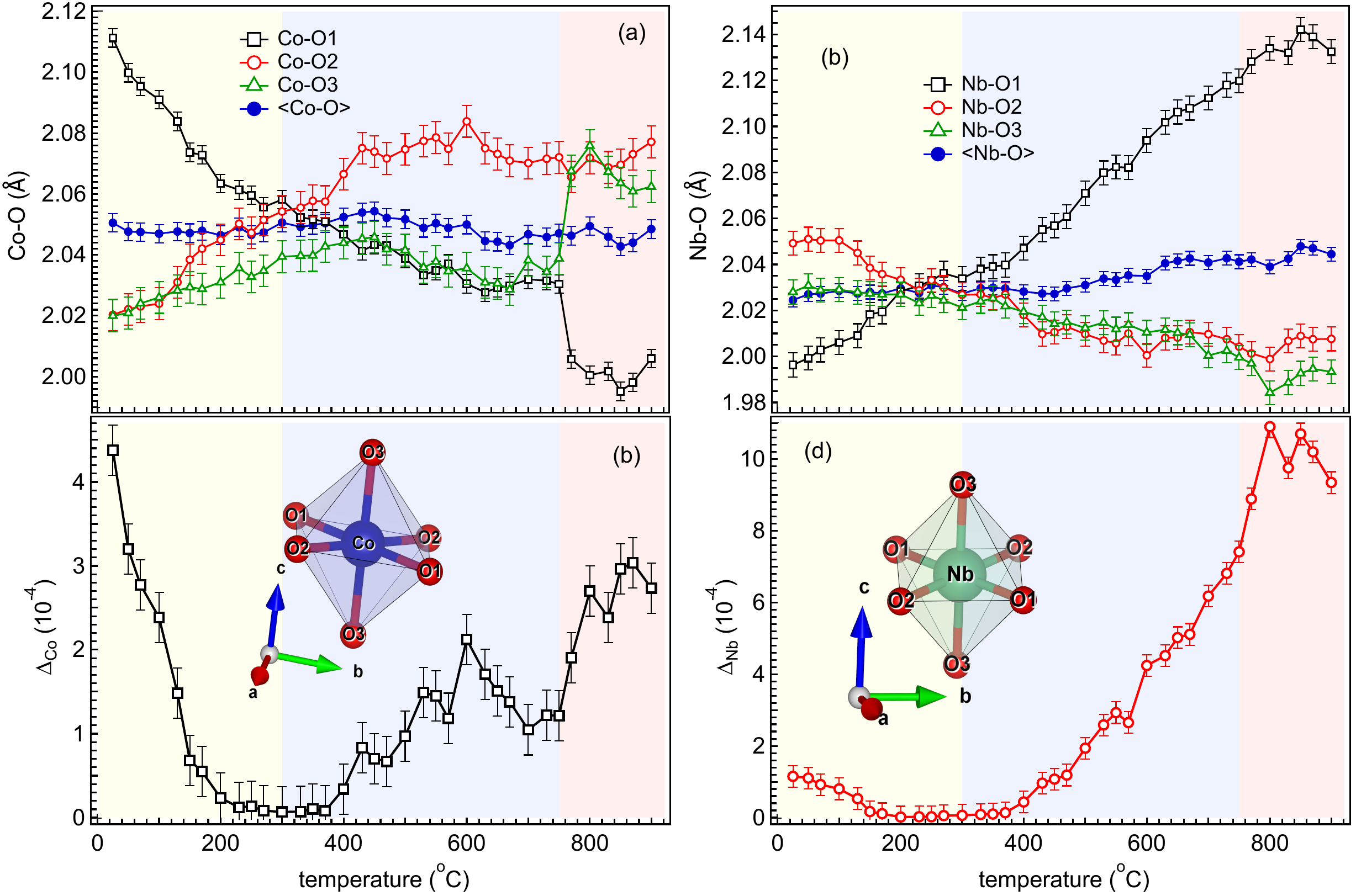}
\caption {Temperature dependence of the different Co/Nb--O bond distances in the (a) CoO$_6$ and (b) NbO$_6$ octahedra, as illustrated in the insets of panels (c) and (d). (c, d) Temperature dependence of the radial distortion, $\Delta$, for the CoO$_6$ and NbO$_6$ octahedra, respectively, calculated using Eq.~\ref{distortion}.} 
\label{Fig10}
\end{figure*}

We now discuss the temperature evolution of the lattice parameters, bond distances, bond angles, and the associated local crystallographic changes. The XRD data recorded using the Cu K$_\alpha$ radiation were employed for this analysis as their longer wavelength provides higher sensitivity to the oxygen sublattice compared to Mo K$_\alpha$ radiation, enabling a more reliable refinement of the oxygen atomic positions and, consequently, a more accurate determination of bond distances and bond angles involving oxygen atoms. Nevertheless, because the x-ray scattering cross section scales approximately with the number of electrons, the positions of lighter atoms such as oxygen are inherently more difficult to determine accurately. Therefore, neutron diffraction measurements would be desirable for a more precise determination of the local octahedral distortions. The temperature dependence of the lattice parameters and unit-cell volume is presented in Fig.~\ref{Fig9}(a--d). Interestingly, the lattice parameters $a$ and $b$ exhibit a crossover around 300~$^\circ$C [Fig.~\ref{Fig9}(a)]. This crossover temperature coincides with the range over which the intensities of the Raman modes and XRD peaks increase upon heating, suggesting a tendency to transform toward higher  symmetry framework. However, none of the Raman or the XRD peaks disappear around this temperature, indicating that the overall crystallographic symmetry remains $P2_1/n$. Instead, the structure evolves toward a less distorted configuration within the same symmetry. Therefore, we designate this temperature as the octahedral-distortion temperature, $T_{\rm OD} = 300\pm 25~^\circ$C. Above $T_{\rm OD}$ the lattice parameters $a$ and $b$ again diverge. This increase in structural distortion results in the deconvolution of Bragg peaks in the XRD patterns above $T_{\rm OD}$ (see Fig.~\ref{Fig4} and Figs.~S5--S8), as discussed earlier.

At higher temperatures ($T \gtrsim 750^\circ$C), a distinct change in the slope of the lattice parameters $a$ and $b$ is observed [Fig. S12 (a, b)], attributed to the $P2_1/n \rightarrow I2/m$ structural phase transition. The cell parameter $c$ also exhibits noticeable changes in slope at both $T_{\rm OD}$ and $T_{\rm PT}$, as shown in Fig.~\ref{Fig9}(b). To enable a direct comparison of the structural parameters across the entire temperature range, all diffraction patterns were refined using the lower-symmetry $P2_1/n$ space group, thereby avoiding possible discontinuities arising solely from the additional symmetry constraints of the $I2/m$ model above $T_{\rm PT}$. The monoclinic angle $\beta$ exhibits a nonmonotonic temperature dependence [Fig.~\ref{Fig9}(c)]. Upon heating, $\beta$ approaches $90^\circ$ and reaches a maximum value near $T_{\rm OD}$. With further increase in temperature, it gradually decreases to $\sim89.80^\circ$ on approaching $T_{\rm PT}$, before increasing again above $T_{\rm PT}$. It is worth mentioning that the monoclinic angle deviates only slightly from $90^\circ$ in the entire temperature range ($\lesssim0.2^\circ$). Owing to the limited $Q$ range of the Cu K$_\alpha$ data ($0.71$--$6.25$ \AA$^{-1}$), the refinement becomes unstable when $\beta$ is allowed to vary freely. Therefore, the diffraction patterns recorded using Mo K$_\alpha$ radiation, which cover a larger $Q$ range of $0.93$--$10.15$ \AA$^{-1}$, were first refined to obtain reliable values of $\beta$ and then the resulting $\beta$ values were subsequently fixed during the refinement of the Cu K$_\alpha$ data.

The unit-cell volume [Fig.~\ref{Fig9}(d)] exhibits relatively weaker anomalies across the transition temperatures compared to the individual lattice parameters [see Fig.~S12 (d)]. This suggests that the expansion along one crystallographic direction at both the transitions is partially compensated by contraction along another direction, and vice versa. The temperature dependence of the unit-cell volume can be approximated by the Grüneisen function given as  \cite{Zhu_PRB_20, Kumar_PRB2_24}

\begin{equation} 
V(T)=V_0 + K_0U(T), 
\label{Gur1}
\end{equation} 

 where $V_0$ is the unit cell volume at 0~K, $K_0$ is the measure of the incompressibility of the sample (a constant), and $U(T)$ is the internal energy, which can be expressed using the Debye function as 

\begin{equation}
U(T) = 9Nk_BT\left(\frac{T}{\theta_D}\right)^3 \int_0^{\theta_D/T} \frac{x^3}{(e^x-1)}dx, 
\label{Gur2}
\end{equation}

where $N$ represents the number of atoms per formula unit (ten in the present case), $k_B$ is the Boltzmann constant, and $\theta_D$ is the Debye temperature. Despite the weak anomalies in the $V(T)$ curve at the transition temperatures, the above equations reproduce the experimental data reasonably well, as shown by the solid curves in Fig.~\ref{Fig9}(d). The best fit to the data in the temperature range $\sim$300--1170~K yields $V_0 = 253.37(2)$~\AA$^3$, $K_0 = 2.20 \times 10^{19}$~\AA$^3$/J, and $\theta_D = 730(13)$~K. The dashed curve in Fig.~\ref{Fig9}(d) represents the extrapolation of the fitted curve down to 0~K.  Importantly, the lattice parameters $a$, $b$, and $c$, as well as the unit-cell volume, vary continuously across both $T_{\mathrm{OD}}$ and $T_{\mathrm{PT}}$ [Figs.~\ref{Fig9}(a, b, and d)], which indicates the second-order-like nature of these transitions. Furthermore, the DSC measurements show no discernible thermal peak associated with latent heat at either transition [see Fig.~S13 and the related discussion in \cite{SI}], supporting this assignment in SrLaCoNbO$_6$.

\section{\noindent ~Discussion }

The Rietveld refinement of the XRD data unambiguously demonstrates that the high-temperature anomaly at $T_{\rm PT}= 750\pm 25~^\circ$C arises from a crystallographic phase transition from $P2_1/n$ to $I2/m$.  The origin of the lower-temperature anomaly near $T_{\rm OD}\approx300 ^\circ$C, however, is less obvious.  Previous spectroscopic and magnetic studies established that Co is predominantly present as high-spin Co$^{2+}$ in SrLaCoNbO$_6$~\cite{Kumar_PRB1_20, Kumar_PRB2_20, Kumar_PRB_22}. Unlike Co$^{3+}$ ($3d^6$), which can exhibit thermally driven low-, intermediate-, and high-spin states~\cite{Radaelli_PRB_02, Kozlenko_PRB_07}, octahedrally coordinated Co$^{2+}$ ($3d^7$) is expected to remain in the high-spin $t_{2g}^{5}e_g^{2}$ configuration in the weak crystal-field environment of SrLaCoNbO$_6$~\cite{Viola_CM_03, Lloret_ICA_08}. Thermal expansion further weakens, rather than strengthens, the crystal-field splitting; therefore, a conventional Co$^{2+}$ spin-state crossover is not expected to be the primary origin of either anomaly. Nevertheless, temperature-dependent x-ray absorption near-edge spectroscopy (XANES), x-ray emission spectroscopy (XES), or high-temperature magnetic measurements would be required to directly examine a possible electronic contribution to the observed phonon renormalizations. Moreover, the absence of discernible differences between the diffraction patterns measured in ambient atmosphere and under high vacuum indicates that oxygen-vacancy-driven structural changes are unlikely to account for the observed transitions. Taken together, these observations support a predominantly structural origin of the observed anomalies.

\begin{figure}  
\centering
\includegraphics[width=1\linewidth]{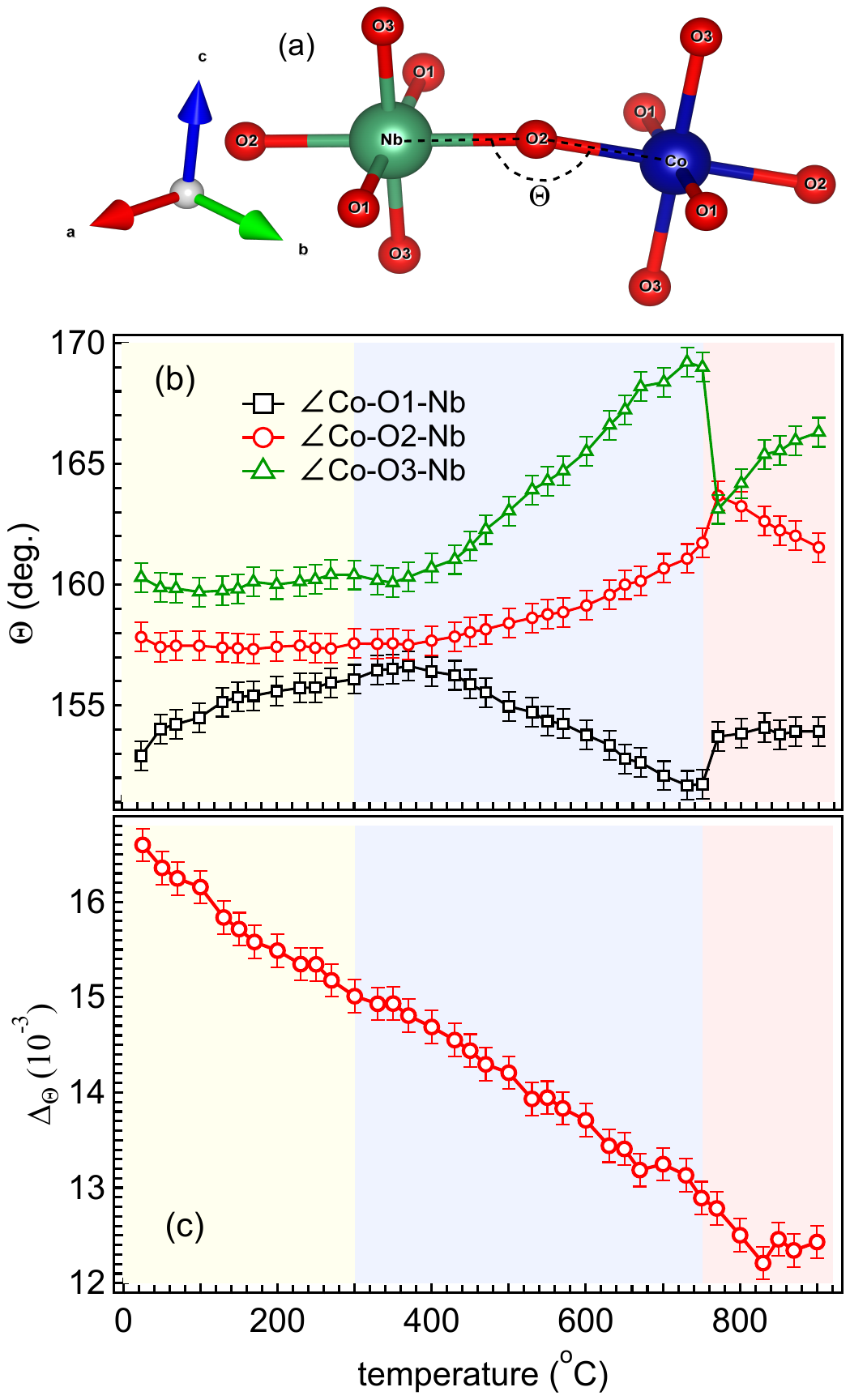}
\caption {(a) Two adjacent CoO$_6$ and NbO$_6$ octahedra in the ordered double perovskite structure of SrLaCoNbO$_6$, shown along the $b$ axis. The Co--O--Nb angle, defined as $\Theta$, represents the octahedral tilting. (b) Temperature dependence of the octahedral tilt angles along all three crystallographic directions. (c) Temperature dependence of the local deviation in the tilt angle $\Theta$ from 180$^\circ$, calculated using Eq.~\ref{distortion2}.} 
\label{Fig11}
\end{figure}

To clarify the nature of low-temperature anomaly, we examine the temperature dependence of the bond distances and bond angles associated with the (Co/Nb)O$_6$ octahedra. In the monoclinic $P2_1/n$ structure, oxygen atoms occupy three inequivalent crystallographic positions, denoted as O1, O2, and O3. The B-site Co and Nb cations are located at the corners of the pseudocubic unit cell and are coordinated by six oxygen atoms, forming CoO$_6$ and NbO$_6$ octahedra. The larger A-site Sr/La cations occupy the center of the unit cell and are coordinated by twelve oxygen atoms. Within the (Co/Nb)O$_6$ octahedra, O1 and O2 are located in the $ab$ plane, whereas O3 lies along the $c$ axis, resulting in two equal B--O bonds associated with each oxygen site. The temperature dependence of the individual Co--O and Nb--O bond distances, together with their average values, is shown in Fig.~\ref{Fig10}(a, b), respectively. Interestingly, the different Co--O and Nb--O bond distances progressively converge upon heating and become nearly identical around $T_{\rm OD}\approx300 ^\circ$C, indicating a substantial reduction in the radial distortion of both CoO$_6$ and NbO$_6$ octahedra. Above $T_{\rm OD}$, the individual Co/Nb-O bond distances again diverge, signaling a re-emergence of the octahedral distortion. More importantly, unlike the conventional Jahn--Teller distortions (observed in Co$^{3+}$-based compounds) typically characterized by differences between apical and equatorial Co--O bond lengths \cite{Kumar_JPCL_22, Fauth_PRB_01, Pradheesh_EPJB_12}, the dominant structural changes in SrLaCoNbO$_6$ occur within the $ab$ plane. In particular, clear crossovers are observed between the Co/Nb--O1 and Co/Nb--O2 bond distances near $T_{\rm OD}$ [Figs.~\ref{Fig10}(a, b)]. These results indicate that the octahedra approach their least distorted configurations near $T_{\rm OD}$, consistent with the simultaneous enhancement of the Raman-mode and XRD-peak intensities observed in this temperature range. In addition, Co--O1 and Co--O3 bond distances exhibit noticeable step-like changes at $T_{\rm PT}\sim750^ \circ$C [Fig. \ref{Fig10}(a)]. This  reflects a substantial reorganization of the octahedral framework accompanying the $P2_1/n \rightarrow I2/m$ phase transition and is consistent with the emergence of new peaks in the XRD patterns above this temperature (Fig.~\ref{Fig8} and Figs.~S10 and S11).

To quantify the octahedral distortion, we calculate the radial distortion parameter, $\Delta$, around both the Co and Nb atoms, defined as \cite{Zhu_PRB_20, Kumar_PRB2_24, Robinson_Sci_71}

\begin{equation}
\Delta = \frac{1}{n} \sum_{i=1}^n \left( \frac{d_n-\langle d \rangle}{\langle d \rangle}\right)^2,
\label{distortion}
\end{equation}

\begin{figure}  
\centering
\includegraphics[width=1\linewidth]{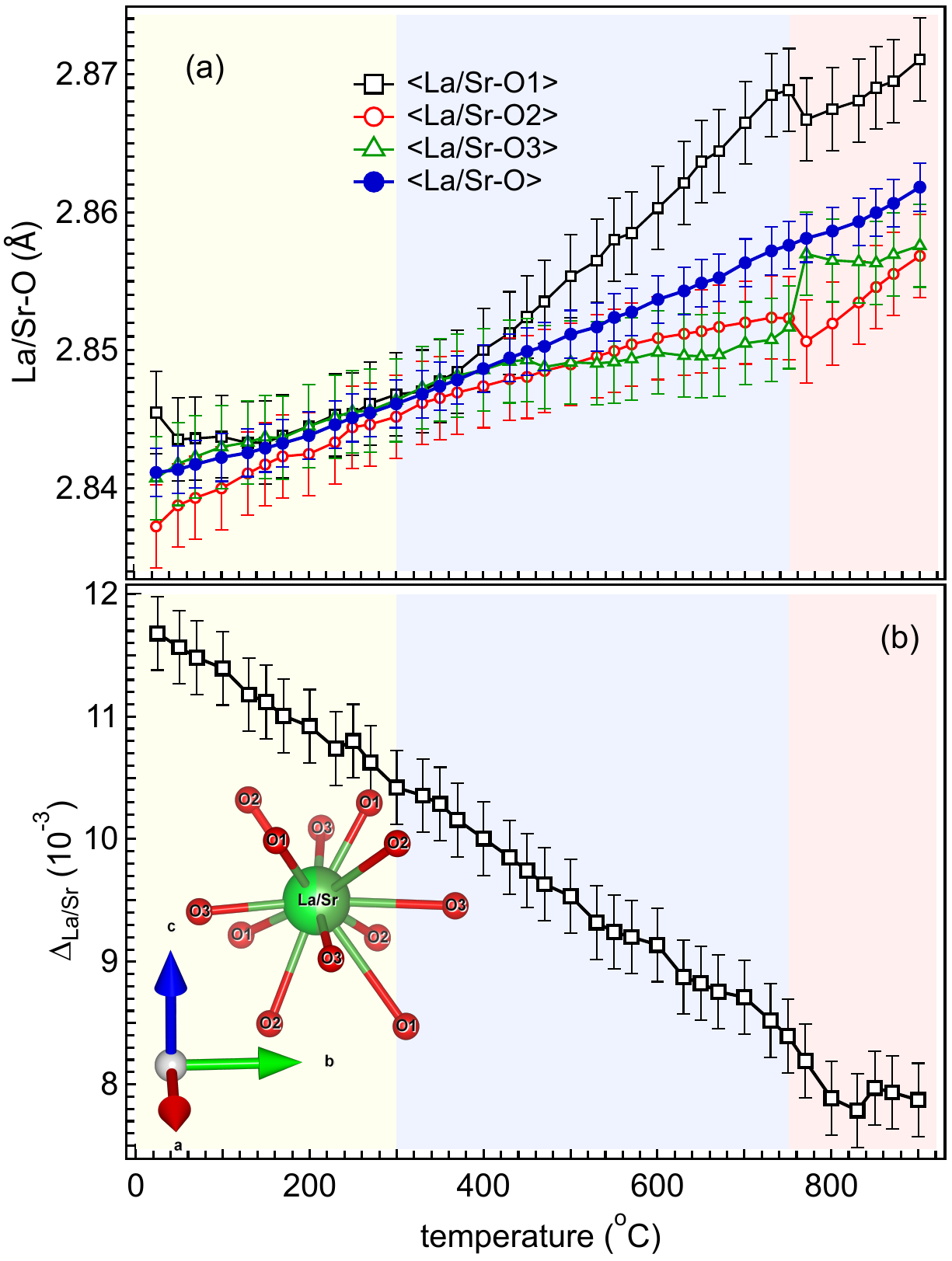}
\caption {(a) Temperature dependence of the average distances between the La/Sr atoms and the three oxygen sites, along with the overall average La/Sr--O bond distance. (b) Local radial distortion, $\Delta$, around the La/Sr atoms, calculated using Eq.~\ref{distortion}. The inset shows a schematic of the La/Sr atom surrounded by twelve oxygen atoms in the $P2_1/n$ symmetry.} 
\label{Fig12}
\end{figure}

where $n$ represents the number of oxygen atoms coordinated to Co/Nb (six in the present case), $d_n$ represents the individual Co/Nb--O bond distances, and $\langle d \rangle$ is the average Co/Nb--O bond distance. The temperature dependence of the distortion parameters around the Co and Nb atoms is presented in Figs.~\ref{Fig10}(c) and (d), respectively. Interestingly, the radial distortion in both the CoO$_6$ and NbO$_6$ octahedra decreases with increasing temperature, approaches nearly zero at $T_{\rm OD}$, and then increases again upon further heating, with a weak anomaly near $T_{\rm PT}$, particularly for the CoO$_6$ octahedra. Further, the angular distortion, quantified by the deviation of the bond angle $\Phi$ (subtended by the two vertically opposite in-plane oxygen atoms at the out-of-plane oxygen atom within the octahedron) from the ideal value of $90^\circ$, closely follows the temperature dependence of the radial distortion parameter $\Delta$ for both CoO$_6$ and NbO$_6$ octahedra (see Fig.~S14 of \cite{SI} and the discussion therein). This correlation is expected because unequal metal--oxygen bond lengths inevitably distort the $\Phi$ angles from their ideal values.

The reversal of the in-plane bond-length anisotropy across $T_{\mathrm{OD}}$ [Co--O1 $>$ Co--O2 and Nb--O1 $<$ Nb--O2 for $T<T_{\mathrm{OD}}$; Co--O1 $<$ Co--O2 and Nb--O1 $>$ Nb--O2 for $T>T_{\mathrm{OD}}$] naturally explains the crossover of the high-energy Raman modes from red to blue shifting at $T_{\mathrm{OD}}$, since these modes are dominated by the symmetric and antisymmetric stretching vibrations of oxygen atoms in the CoO$_6$ and NbO$_6$ octahedra \cite{Iliev_PRB_07, Iliev_PRB_98}. In contrast, no anomaly is observed in the position of the low-energy bending and translational modes near $T_{\mathrm{OD}}$. To understand this, we first examine the octahedral tilting through the Co--O--Nb bond angle, $\Theta$ [Fig.~\ref{Fig11}(a)]. The temperature evolution of the bending angle $\Theta$ along the three crystallographic directions is shown in Fig.~\ref{Fig11}(b). Notably, the three bond angles approach each other near $T_{\rm OD}$. Above $T_{\mathrm{OD}}$, they diverge, where the Co--O1--Nb angle decreases while the Co--O2--Nb and Co--O3--Nb angles increase up to $T_{\mathrm{PT}}$, followed by a distinct change  in their behavior near $T_{\rm PT}$.

The average deviation in the bending angle, $\Delta_\Theta$, is calculated from 180$^\circ$ using the following equation \cite{Robinson_Sci_71}

\begin{equation}
\Delta_\Theta = \frac{1}{n} \sum_{i=1}^n \left( \frac{\Theta_n-180}{180}\right)^2,
\label{distortion2}
\end{equation}

where $n$ represents the number of Co--O--Nb bond angles (three in the present case) and $\Theta_n$ denotes the individual Co--O--Nb bond angles. The temperature dependence of $\Delta_\Theta$ is shown in Fig.~\ref{Fig11}(c), which interestingly exhibits a monotonic decrease with increasing temperature with no distinct anomaly at $T_{\mathrm{OD}}$. This indicates that the Co--O--Nb bonds progressively straighten toward the ideal $180^\circ$ geometry upon heating, with no discernible change  across $T_{\mathrm{OD}}$. However, when $\Delta_\Theta$ is calculated from the average values of the Co--O--Nb bond angles, it shows the distinct anomalies at both $T_{\mathrm{OD}}$ and  $T_{\rm PT}$ as shown in Fig. S15 of \cite{SI}.

Finally, we examine the radial distortion around the $A$-site (La/Sr) atoms, which are coordinated by twelve oxygen atoms, resulting in twelve inequivalent La/Sr--O bond lengths, where each oxygen atom participates in four nonequivalent La/Sr--O bonds. The temperature dependence of these bond lengths (Fig.~S16) reveals changes near both $T_{\rm OD}$ and $T_{\rm PT}$, with the latter being more pronounced. The average La/Sr--O bond distances corresponding to the three oxygen sites are shown in Fig.~\ref{Fig12}(a). All three average La/Sr--O bond distances increase with temperature; however, they approach each other around $T_{\rm OD}$ and exhibit a distinct anomaly near $T_{\rm PT}$. The local distortion around the La/Sr atoms, calculated using the individual La/Sr--O bond distances (Fig.~S16) and Eq.~\ref{distortion}, is plotted in Fig.~\ref{Fig12}(b).  $\Delta_{\rm La/Sr}$ decreases monotonically with temperature without any distinct anomaly across $T_{\rm OD}$. In contrast, $\Delta_{\rm La/Sr}$ evaluated using the average La/Sr--O bond distances exhibits distinct anomalies at both $T_{\rm OD}$ and $T_{\rm PT}$ (Fig.~S17). The monotonic decrease of both $\Delta_\Theta$ [Fig.~\ref{Fig11}(c)] and $\Delta_{\rm La/Sr}$ [Fig.~\ref{Fig12}(b)] across $T_{\rm OD}$ is consistent with the absence of any discernible anomaly in the frequencies of the low-energy Raman modes P3, P4, P7, and P8 near this temperature. This suggests that these modes are primarily coupled to octahedral tilting and the motion of $A$-site cations \cite{Andrews_DT_15, Ayala_JAP_07}. Nevertheless, detailed theoretical calculations would be valuable for establishing the microscopic origin of the individual Raman modes and their direct correlation with the corresponding structural distortions.

\section{\noindent ~Conclusion}

We have investigated the high-temperature structural evolution of the B-site ordered double perovskite SrLaCoNbO$_6$ through combined temperature-dependent Raman spectroscopy and laboratory high-flux powder XRD in the temperature range from  25 to  900~$^\circ$C. Two distinct structural anomalies are identified at $T_\mathrm{OD} = 300\pm 25~^\circ$C and $T_\mathrm{PT} = 750\pm 25~^\circ$C with fundamentally different microscopic origins. The low-temperature anomaly corresponds to an isostructural octahedral distortion within the $P2_1/n$ phase, wherein the radial and angular distortion parameters of the CoO$_6$ and NbO$_6$ octahedra decrease to near-zero values before reversing upon further heating---a crossover driven by an inversion of  the in-plane Co--O and Nb--O bond-length hierarchy and directly  reflected in the anomalous behavior of the high-frequency Raman stretching modes. The high-temperature anomaly, in contrast,  corresponds to a symmetry-breaking phase transition from $P2_1/n$ to  $I2/m$, confirmed by the suppression of characteristic  XRD reflections and abrupt changes in the Raman phonon parameters.  Overall, this work establishes an effective pathway to disentangle isostructural octahedral distortions from symmetry-breaking phase transitions in complex double perovskite oxides through combined Raman spectroscopy, laboratory-based XRD, and the subsequent quantitative analysis of the distortion parameters, without relying on extensive synchrotron-based measurements.

\begin{figure*}  
\centering
\includegraphics[width=0.9\linewidth]{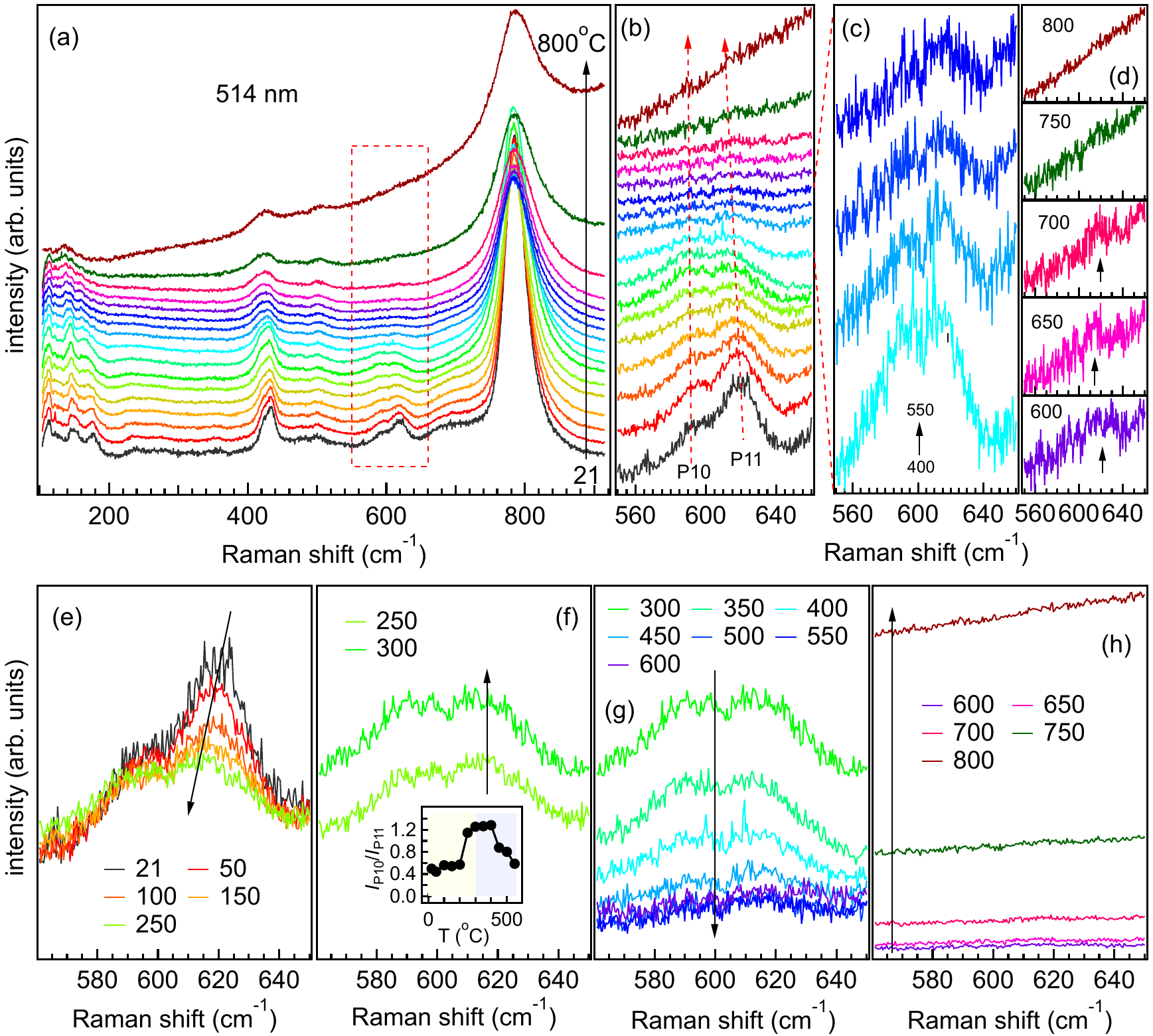}
\caption {(a) Raman spectra of SrLaCoNbO$_6$ recorded from 21~$^\circ$C to 800~$^\circ$C using a 514~nm excitation source. For clarity, the spectra have been vertically offset. (b--d) Enlarged views of modes P10 and P11 over the entire temperature range, 400--550~$^\circ$C, and 600--800~$^\circ$C, respectively. (e--h) Temperature evolution of modes P10 and P11 without vertical offset. The arrows indicate the direction of increasing temperature. The intensity scale is identical for panels (e--g), whereas a different scale is used for panel (h) for clarity. The inset of panel (f) shows the temperature dependence of the integrated intensity ratio, $I_{\mathrm{P10}}/I_{\mathrm{P11}}$.} 
\label{Fig13}
\end{figure*}

  \section{\noindent ~Acknowledgments}

The temperature-dependent XRD and DSC measurements, and the data analysis were performed at Ames National Laboratory, USA (A. K. and Y. M.) and were supported by the Division of Materials Science and Engineering of the Office of Basic Energy Sciences, Office of Science of the U.S. Department of Energy (DOE). Ames National Laboratory is operated for the U.S. DOE by Iowa State University of Science and Technology under Contract No. DE-AC02-07CH11358. The sample preparation, FESEM, and HRTEM measurements were performed at IIT Delhi. The high-temperature Raman spectroscopy measurements were carried out at UNSW. RSD gratefully acknowledges the Department of Science and Technology (DST), India, for support through Indo-Australia Early and Mid-career Researchers fellowship (IA/INDOAUST/F-19/2017/1887). We acknowledge support from SERB-DST through
a core research grant (No. CRG/2020/003436). CU acknowledges the support of the Australian Research Council through the Discovery Grant No. DP160100545.

\section*{APPENDIX: Additional Raman spectroscopy data}

Figure \ref{Fig13}(a) shows the Raman spectra of SrLaCoNbO$_6$ recorded between 21~$^\circ$C and 800~$^\circ$C using a 514~nm excitation source. For clarity, each spectrum has been vertically shifted. Figures \ref{Fig13}(b--d) display enlarged views of modes P10 and P11 [highlighted by the dashed rectangle in panel (a)] over the temperature ranges of 21--800~$^\circ$C, 400--550~$^\circ$C, and 600--800~$^\circ$C, respectively. It can be observed that both modes disappear above $\sim$700~$^\circ$C, as indicated by the vertical arrows in panel (d). Figures \ref{Fig13}(e--h) show enlarged views of the unshifted P10 and P11 modes in different temperature intervals. The arrows indicate the direction of increasing temperature. It is evident that the intensities of both modes decrease with increasing temperature up to $\sim$250~$^\circ$C [panel (e)], followed by an increase around 300~$^\circ$C [panel (f)]. Subsequently, the intensities decrease again with further heating up to about 600~$^\circ$C [panel (g)] before increasing once more at higher temperatures [panel (h)]. Note that the intensity scale is identical for panels (e--g), whereas a different scale is used for panel (h) to improve visibility. The inset of panel (f) presents the temperature dependence of the integrated intensity ratio of modes P10 and P11, $I_{\mathrm{P10}}/I_{\mathrm{P11}}$. This demonstrates that mode P10 gains spectral weight relative to P11 around 300~$^\circ$C, followed by a gradual reduction in the intensity ratio with further increase in temperature.

\end{document}